\documentstyle[12pt]{article}
\begin{document}
 \title{Angular and energy dependence of $(e,e^{\prime})$ \\ 
 cross sections for orbital 1$^+$ excitations}
\author{R. Nojarov\thanks{ 
 Permanent address: Institute for Nuclear Research 
 and Nuclear Energy, Bulgarian Aca\-de\-my of Sciences, 
 BG-1784, Sofia, Bulgaria. E-mail: roland.nojarov@uni-tuebingen.de}, 
 Amand Faessler\thanks{E-mail: amand.faessler@uni-tuebingen.de} 
 and M.~Dingfelder\thanks{E-mail: michael.dingfelder@uni-tuebingen.de}
   \\ 
Institut f\"ur Theoretische Physik, Universit\"at T\"ubingen, \\
Auf der Morgenstelle 14, D-72076 T\"ubingen, Germany}
\date{\today}
\maketitle
 \begin{abstract}
The main features of the $(e,e^{\prime})$ cross sections of low-lying
orbital excitations with $K^{\pi} = 1^+$ in heavy deformed nuclei are
studied in RPA on the example of $^{156}$Gd. The dependence of the
DWBA E2 and M1 cross sections on the scattering angle $0^{\circ} <
\theta < 180 ^{\circ}$ and incident electron energy $E_i < 210$ MeV
is analyzed in PWBA. The cross section is larger for M1 than for E2
transitions at any angle  if $E_i < 30$ MeV.  The longitudinal
(Coulomb) C2 excitation dominates the E2 response for  $5^{\circ} <
\theta < 170 ^{\circ}$. Only transverse M1 and E2 excitations compete
for $\theta > 175 ^{\circ}$ and the former one is dominant for $q <
1.2$ fm$^{-1}$. The M1 response is almost purely orbital up to $q =
1.4$ fm$^{-1}$ even in backward scattering.  Qualitative PWBA
estimates based on the $q$-dependence of the form factors alone are
not able to predict some important features  of the $(e,e^{\prime})$
cross sections stemming from the strong magnetic and orbital
character of the studied 1$^+$ excitations.  The expectation for M1
over E2 dominance in backward scattering should not be extended to
higher momentum transfers and incident energies. 
\end{abstract}

\newpage

\section{Introduction}

The backward inelastic electron scattering has established itself 
in the last 30 years as one of the main tools for the experimental
study of nuclear magnetic excitations. The attention was focused in
the past mainly on spherical nuclei, as it can be seen, e.g., from
the review articles \cite{fagg,ramfagg}.  The detailed study of
low-lying magnetic dipole (M1) excitations in heavy deformed nuclei
started with  $(e,e^{\prime})$ experiments at the linear
accelerator in Darmstadt \cite{bohle84}, reviewed recently in
\cite{achimrev}. Most of the data were collected at a backward
scattering angle $\theta = 165 ^{\circ}$, where one expects a strong
damping of electric excitations.

This can be understood from qualitative considerations
\cite{bar62,bar63} in the plane wave Born approximation (PWBA).  The
$(e,e^{\prime})$ cross section can be decomposed in this case into
longitudinal (Coulomb) and transverse (electric and magnetic) terms,
multiplied by corresponding kinematic factors. The longitudinal
kinematic factor $V_{\ell} (\theta )$ vanishes for $\theta =
180^{\circ}$ and only transverse multipoles are excited in backward
scattering \cite{bar62}. Among those with the same multipolarity, the
magnetic excitation is dominant  over the electric transverse one
\cite{bar63}. This qualitative estimate is obtained in the long-wave
limit  $qR \ll 1$, or small momentum transfer $q$ in comparison with
the radius  $R$ of the target nucleus.  Magnetic dominance was found
at backward angle also in \cite{fagg} by assuming a purely spin-flip
transition.

The more realistic distorted wave Born approximation (DWBA)
\cite{uber} leads to important corrections especially in heavy nuclei
with a strong Coulomb field (a large charge number $Z$). But this
approach is not suitable for simple qualitative predictions, since
the longitudinal and transverse electric contributions to the
$(e,e^{\prime})$ cross section interfere with each other in DWBA.
Moreover, their separation would be meaningless when the longitudinal
and transverse electric contributions are related in the DWBA
formalism \cite{heis,heis81} through the continuity equation.
 
In contrast to the above discussed common expectations based on the
PWBA, we have found recently \cite{nfd94,dnf95} E2 contributions to
the cross sections of low-lying orbital M1 excitations measured at
$\theta = 165 ^{\circ}$.  Such admixtures take place in heavy
deformed nuclei where the M1 transition with $I^{\pi}K = 1^+1$ is
accompanied by a closely lying E2 transition with $I^{\pi}K = 2^+1$
to the first member of the rotational band built on the intrinsic
$K^{\pi} = 1^+$ excitation.  The large moment of inertia of these
nuclei leads to a small separation between the two excitation
energies, comparable with the energy resolution of the
$(e,e^{\prime})$ experiments.  The accompanying E2 transition
provides an important contribution to the measured cross sections
($\theta = 165 ^{\circ}$) for an effective momentum transfer $0.6 <
q_{\rm eff} < 0.9$ fm$^{-1}$ corresponding to incident electron
energies in the range $40 < E_i < 70$ MeV \cite{nfd94,dnf95}.

The M1 excitations in spherical nuclei are mainly of spin-flip type, 
which is expected to be favoured in backward electron scattering
\cite{ramfagg}. In contrast, we are considering here a qualitatively
different type of predominantly orbital M1 excitations. They appear
only in deformed nuclei \cite{noj95} and their experimental study
through backward $(e,e^{\prime})$ scattering started much later
\cite{bohle84,achimrev}.

We are going to examine here in more detail some of the well-known
qualitative PWBA predictions for the interplay of magnetic and
electric excitations, paying attention also to the approximations
involved.  These predictions ignore the nuclear dynamics because they
are made for the general case of unknown transition densities. The
latter could exhibit however some important peculiarities, typical
for the considered nuclear excitations. The qualitative PWBA
estimates will be compared here to DWBA and PWBA cross sections
obtained from realistic RPA transition densities in order to
understand to what extent one could use the general PWBA predictions
for qualitative interpretation of realistic microscopic results.

We shall study to this purpose the dependence of our theoretical
DWBA $(e,e^{\prime})$ cross sections on the electron incident energy
and momentum transfer in the whole possible range of scattering
angles $0 ^{\circ} < \theta < 180^{\circ}$.  Cross sections of 1$^+$
excitations in deformed nuclei have been presented and discussed
until now almost exclusively for $\theta = 165 ^{\circ}$ in order to
compare them with the corresponding experimental data measured at
this angle. Only the cross section of the strongest (low-lying
orbital) M1 excitation is known experimentally in each one of the
several heavy deformed nuclei studied until now.  We choose here
$^{156}$Gd as an example, because this is the only nucleus, where the
accompanying E2 transition was identified experimentally 
\cite{bohle85}. 

We shall present and discuss in the next section the dependence of E2
and M1 cross sections of this typical 1$^+$ excitation on the
incident electron energy and momentum transfer in the whole range of
scattering angles. The results are analyzed qualitatively in PWBA for
small and large momentum transfer in sections 3 and 4, respectively.
The conclusions are summarized in sect. 5 and expressions for E2
transition densities are given in the Appendix.

\section{Angular and energy dependence of E2 and M1 $(\lowercase{e}, 
   \lowercase{e} ^{\prime})$  cross sections}

We describe intrinsic excitations with $K^{\pi} =1^+$ in deformed
nuclei within the quasiparticle random phase approximation (QRPA, or
shorter RPA) using a model hamiltonian specified in \cite{nfd95}. It
contains a quasiparticle (q.p.) mean field, given by a deformed
Woods-Saxon potential plus pairing. The separable residual
interaction consists of a spin-spin force and a quadrupole-type 
interaction, derived self-consistently from the deformation of the
mean field. The isoscalar coupling constant of the latter
interaction, determined microscopically, ensures the rotational
invariance of the model hamiltonian, which is violated by the mean
field alone. 

The choice of the parameters of the model hamiltonian is explained in
more detail elsewhere \cite{nfd95}. In contrast to many other
microscopic calculations, energies of single-particle levels are not
shifted but taken exactly as provided by the deformed Woods-Saxon
potential. Expressions for M1 and E2 transition probabilities in terms
of the RPA amplitudes are given in \cite{nofa88}. The large
single-particle basis allows us to work without effective charges
$\varepsilon _{\rm eff}$ when calculating $B(E2; 0^+0 \to 2^+1)$
values and E2 transition densities.  The $B(M1; 0^+0 \to 1^+1)$
values and M1 transition densities are obtained with bare orbital and
effective spin gyromagnetic factors.  The latter account for the
renormalization of the spin operator by a factor of 0.7, i.e.,
\begin{equation}
 \varepsilon ^n =0, \; \varepsilon ^p =1, \; \; g^s_j = 0.7 g^s_j 
 ({\rm free}), \quad
g^s_n ({\rm free}) = -3.8263, \; \;  g^s_p ({\rm free}) = 5.5858. 
\label{grat} 
\end{equation}

The  strongest M1 transition found experimentally in $^{156}$Gd has 
an excitation energy $E_x = 3.07$ MeV and a reduced magnetic
transition probability $B$(M1) = 1.30 $\pm 0.20 \; \mu_N^2$
\cite{bohle86}. It is easily identified with the strongest
theoretical M1 transition at 2.90 MeV with $B$(M1) = 1.24 $\mu_N^2$.
Our theoretical reduced probability for the accompanying E2
transition,  $B$(E2) = 42 e$^2$fm$^4$, agrees also well with the
corresponding experimental value  40 $\pm 6$ e$^2$fm$^4$
\cite{bohle85}. The $B$(E2) values of $K^{\pi} =1^+$ excitations are
small because the main collectivity of the E2 strength of this mode
is concentrated in the spurious rotational state, i.e. in the lowest
E2 transition of the ground state rotational band.

The $(e,e^{\prime})$ cross sections are, however, much more sensitive
to the details of the RPA wave functions than transition
probabilities or excitation energies. We obtain the transition
densities, necessary for the calculation of DWBA cross sections, as
reduced matrix elements of the corresponding multipole density
operators between the nuclear ground state and the $\nu$-th RPA
excitation with $I^{\pi}K = L^+1 \; (L=1,2)$:
\begin{eqnarray}
\rho _2 ^{\nu} (r) = \langle 2^+1, \nu \parallel \hat{\rho} 
^{\dag}_2 (r) \parallel 0 \rangle , 
\label{rhoden} \\
{\cal J} _{LL^{\prime}} ^{\nu} (r) = i \langle L^+1, \nu \parallel 
\hat{\cal J} ^{\dag} _{LL^{\prime}} (r) \parallel 0 \rangle, 
\quad 
 {\cal J}^{\nu} _{LL^{\prime}} (r) =  {\cal J}^{\nu ,C}
_{LL^{\prime}}  (r) + {\cal J}^{\nu ,S} _{LL^{\prime}} (r), 
\label{jden} 
\end{eqnarray}
where $\hat{\rho} _2 (r)$ and $\hat{\cal J} _{LL^{\prime}} (r)$ are 
longitudinal charge (C2) and transverse current (T$L$) density
operators, respectively. The latter reduce to $\hat{\cal J} _{11}
(r)$ for M1 excitations ($L = L^{\prime} =1$) and $\hat{\cal J}
_{2L^{\prime}} (r)$ for E2 excitations ($L = 2, \; L^{\prime} =1,3$).
The transverse (current) transition densities $ {\cal J}^{\nu}
_{LL^{\prime}} (r)$ are written in (\ref{jden}) as a sum of two terms
originating from the proton convection current density ${\hat {\bf
j}} ^{\rm C} ({\bf r})$ and the magnetization (or spin) current
density ${\hat {\bf j}} ^{\rm S} ({\bf r})$ of protons and neutrons.

The current density operators and expressions for the M1 transition
densities in RPA can be found in \cite{fanota}. Expressions for the
E2 transition densities are given in the Appendix.  The
$(e,e^{\prime})$ cross sections are calculated numerically from the
transition densities (\ref{rhoden}), (\ref{jden}), using the DWBA
formalism described in \cite{heis,heis81}.

The theoretical DWBA cross sections of the above described low-lying
orbital 1$^+$ excitation in $^{156}$Gd are displayed in Figs.
\ref{fig1} and \ref{fig2}. The M1 and E2 $(e,e^{\prime})$ cross
sections are plotted in Fig. \ref{fig1} versus the incident electron
energy in the range $10 \le E_i \le 210$ MeV and scattering angles
$5^{\circ} \le \theta \le 175^{\circ}$. 

Since we are interested here mainly in qualitative conclusions, the
agreement with experiment is relevant to the extent it could promote
the confidence in the theoretical cross sections beyond the
experimentally studied scattering angle of 165$^{\circ}$.
Two-dimensional sections  of the three-dimensional plots in Fig.
\ref{fig1}, corresponding to a fixed angle $\theta = 165^{\circ}$ and
$E_i < 100$ MeV, were plotted together in \cite{dnf95} and compared
with experimental data \cite{bohle84}, taken at this angle.  The E2
excitation provides important contributions to the measured M1 cross
section in the region $40 < E_i < 70$ MeV and leads to a very good
agreement with experiment after being taken into account.

These features can be seen also by comparing the shaded areas
($\theta = 175^{\circ}$) of the M1 and E2 cross sections in Fig.
\ref{fig1}.  For small incident energies $E_i$ the E2 cross section
is two orders of magnitude smaller than the M1 cross section but both
are comparable at higher $E_i$.  At 175$^{\circ}$ the two cross
sections are comparable for $120 < E_i < 200$ MeV and the E2 response
is dominant for $E_i > 200$ MeV.

We are going to analyze qualitatively our microscopic DWBA cross
sections using PWBA, where the cross section can be decomposed into a
sum of longitudinal and transverse terms \cite{uber,heis}:
\begin{eqnarray}
\biggl ( {d \sigma \over d \Omega} \biggr )_{\rm PWBA} = \biggl ( 
{Z e^2 \over E_i} \biggr ) ^2 \Bigl \{ V_{\ell} \vert F^C_L (q) 
\vert ^2 
+ V_t \bigl [ \vert F^E_L (q) \vert ^2  + \vert F^M_L (q) \vert ^2 
\bigr ] \Bigr \},  \label{pwba} \\ 
V_{\ell} = { {\rm cos}^2 (\theta /2) \over 4 {\rm sin}^4 
(\theta /2)}, \; \; 
V_t = { 1 +  {\rm sin}^2 (\theta /2) \over 8 {\rm sin}^4 
(\theta /2)},  \quad  
V_T = V_t/V_{\ell} = {1 \over 2} + \tan ^2 \left ( {\theta /2}  
 \right ). \label{vlt}
\end{eqnarray}
$Z$ and $e = \sqrt{\alpha \hbar c}$ are the nuclear proton number and
the unit charge, respectively. The factor  $4 \pi /Z^2$ from
\cite{uber,heis} is incorporated in our definitions \cite{fanota} of 
the form factors $\vert F_L(q) \vert ^2$ (\ref{pwba}), coinciding
with those of \cite{bar63}. Only the term $F^M_1(q)$  survives in
(\ref{pwba}) in the case of $M1$ transitions, while the $E2$
excitations are specified by longitudinal (Coulomb) $F^C_2(q)$ and
transverse electric $F^E_2(q)$ contributions. The transferred
momentum has in our notations the inverse dimension of length, $q$
(fm$^{-1}$).

The PWBA formalism is used in this paper for two different purposes.
First, in addition to the DWBA results, cross sections are
calculated in some cases also in PWBA with the same RPA transition
densities in order to separate the longitudinal C2 from the
transverse E2 response and compare them with the transverse M1 cross
sections.  This separation is not possible in DWBA. Hence, the
theoretical PWBA cross sections are used only as a tool for studying
different constituents of the cross sections and their origin.
Secondly,  DWBA and PWBA cross sections, both obtained from realistic
RPA transition densities, are compared to qualitative PWBA
predictions which ignore completely the nuclear dymanics. These
estimates are based on general considerations of the scattering
kinematics and the momentum dependence of the unknown form factors.
On the other hand, the latter are fully specified in our formalism by
the microscopic transition densities (\ref{rhoden}), (\ref{jden}).

It is seen from Fig. \ref{fig1} that for small incident energies the
$M1$ cross section remains almost constant when going from backward
to forward angles.  It starts to increase for $\theta < 90^{\circ}$
but, contrary to what should be expected from the diverging
transverse kinematic factor $V_t$ (\ref{vlt}), the $M1$ cross section
does not diverge in the limit $\theta \to 0^{\circ}$. The expressions
(\ref{vlt}), derived in the relativistic limit (neglected electron
rest mass), must be corrected for extreme forward or backward angles
and acquire finite values, as we shall see in the next section.  The
DWBA cross section of the accompanying $E2$ transition, displayed in
Fig.\ \ref{fig1}, decreases substantially in the backward direction
(large $\theta$) for small incident energies.  This is an indication
for a dominant Coulomb response which is damped by the longitudinal
factor $V_{\ell}$ at backward angles.

DWBA cross sections are plotted usually \cite{uber,heis81} versus
the effective momentum transfer $q_{\rm eff}$, 
\begin{equation}
q_{\rm eff} = q \biggl [ 1 + {3 \over 2} { Z \alpha \hbar c \over 
E_i R_{\rm eq} } \biggr ], \quad q = { 2 \over \hbar c} \sqrt{E_i E_f}
{\rm sin} (\theta /2) = {2 E_i \over \hbar c} {\rm sin} (\theta /2) 
\sqrt{1 - (E_x /E_i)},  \label{qeff}
\end{equation}
in order to be compared with PWBA cross sections plotted versus $q$.
In the above expression  $E_f = E_i - E_x$ is the energy of the
outgoing electron. The definition (\ref{qeff}) coincides with the
formula for $q_{\rm eff}$ in \cite{fanota} used to plot the
experimental cross sections when a radius constant $r_{\rm eq} =
1.12$ fm \cite{heis} is assumed in (\ref{qeff}) for the equivalent
nuclear radius $R_{\rm eq} = r_{\rm eq} A^{1/3}$.

The form factors $\vert F_L(q) \vert ^2$ (\ref{pwba}) are Fourier
transforms of the transition densities (\ref{rhoden}), (\ref{jden})
with Bessel functions $j_L(qr)$ \cite{uber,heis81}. Therefore, the
explicit dependence of the theoretical M1 and E2 cross sections on
the momentum transfer is also of prime interest. It is plotted in
Fig. \ref{fig2} versus $q_{\rm eff}$ (\ref{qeff}) and the scattering
angle $\theta$.  The indented edges in Fig. \ref{fig2} result simply
from the adopted finite grid of mesh points. The graph cut-off at the
edge corresponds to the maximal incident energy $E_i = 200$ MeV
included in calculations.

Let us rewrite the kinematic factors in a form that is more
appropriate for a constant $q$. We neglect the small term $E_x/E_i$
in (\ref{qeff}) and express $E_i^{-2}$ in (\ref{pwba}) through $q$:
\begin{eqnarray}
q =  {2 E_i \over \hbar c} \sin (\theta /2), 
\label{qe} \\
{ V_{\ell} \over E_i^2 } = { 1 \over q^2} \Bigl [ \sin ^{-2} 
(\theta /2) -1 \Bigr ], \quad  
{ V_t \over E_i^2 } = { 1 \over 2 q^2} \Bigl [ \sin ^{-2} 
(\theta /2) +1 \Bigr], \nonumber \\ 
d \sigma_{\ell} (\theta ) = d \sigma_{\ell} (90^{\circ}) \cot ^2 
(\theta /2), \quad 
d \sigma_t (\theta ) = d \sigma_t (90^{\circ}) { 1 \over 3} \biggl [
{1 \over  \sin ^2 (\theta /2) } +1 \biggr ]. 
 \label{vltq}
\end{eqnarray} 
These expressions give the angular dependence of longitudinal and
transverse cross sections for a fixed momentum transfer $q$.  
It is seen from (\ref{vltq}) that for a given $q$ the transverse
cross section decreases only 33\% between 90$^{\circ}$ and
180$^{\circ}$.  In contrast, the longitudinal cross section decreases
in comparison with its value at 90$^{\circ}$: 58 times at
165$^{\circ}$, where most experiments were done, and 5835 times at
the largest accessible backward angle of 178.5$^{\circ}$
\cite{petra}.

Thus, the almost constant M1 cross section in Fig. \ref{fig2} for a
given momentum transfer is a typical behaviour of transverse
excitations, while the steep decrease of the E2 cross section towards
backward angles is an indication for a predominant longitudinal
(Coulomb) component. 

\section{PWBA analysis for small momentum transfer}

The low-$q$ limit $qR < 1$ is reached in the considered heavy nuclei
with $R \approx 6.5$ fm for very small momentum transfer $q < 0.15$
fm$^{-1}$. The PWBA cross section (\ref{pwba}) is obtained in the
relativistic limit, i.e. the electron rest mass $m_e c^2 = 0.511$ MeV
is neglected in comparison with the incident electron energy $E_i$
and a small excitation energy $E_x \ll q\hbar c$ is assumed in
comparison with the momentum transfer. The former condition is always
satisfied in our calculations because $E_i > 10$ MeV. The latter
condition is violated at small angles, where one has to use the more
accurate expressions provided by Eqs. (4-16c) and (6-38) of ref.
\cite{uber}:
\begin{eqnarray}
V_T  = { q^2 \over \Delta ^2} \biggl [ {1 \over 2} + { q^2 \over
\Delta ^2} \tan ^2    \left ( {\theta /2} \right ) \biggr ], 
\quad \Delta ^2 = q^2 - \bigl ( E_x / \hbar c \bigr )^2 , 
\label{vtd} \\ 
\mbox{for } \theta = 0^{\circ}: \; \; V_{\ell} = 4 \biggl ( {E_i 
\over E_x} \biggr ) ^4, \; \; V_t = { 2 E_i^6 \over E_x^4 
(m_ec^2)^2},  \; \;  V_T  = {1\over 2} \biggl ( {E_i \over m_e
c^2 } \biggr ) ^2 .
\label{vtzero}
\end{eqnarray}
The electron rest mass is still neglected in (\ref{vtd}), which can
not be used for $\theta = 0$ since the four-momentum is $\Delta = 0$
and the transverse factor diverges. After taking the electron mass
into account, one obtains  $\Delta ^2 (\theta = 0) = (m_ec^2 E_x /E_i
)^2$ and the finite values (\ref{vtzero}) in the forward direction.
The transverse cross section is strongly dominant for any realistic
incident energy $E_i \gg E_x > m_ec^2$, as seen from (\ref{vtzero}).
The limit (\ref{vtzero}) is however not necessary even for a very
small but different from zero angle (a fraction of the degree)
because the expression (\ref{vtd}) provides already a very good
numerical accuracy \cite{willey,uber}.  The smallest angle displayed
in the three-dimentional graphs is $\theta = 5^{\circ}$.

It is seen from (\ref{vtd}) that $V_T$ is a very large number for
small angles where $q$ is close to the photon point $E_x / \hbar c$
and $\Delta$ is small. One can understand in this way the behaviour
of the M1 and E2 cross sections at small scattering angles, exhibited
in Fig. \ref{fig1}. The M1 cross section is dominant in the forward
direction, in line with the general expectations \cite{raph66,uber}
for suppression of the longitudinal response throught its small
kinematic factor at small angles.  But this is no more true for
higher momentum transfer.  The factor $V_T$ (\ref{vtd}) that damps
the longitudinal (Coulomb) excitation is large at $\theta =
5^{\circ}$ only for small momentum transfer.  However, Eq.
(\ref{vtd}) reduces already to the simpler estimate (\ref{vlt}) $V_T
\approx 0.5$  for $E_i = 200$ MeV. The Coulomb response is favoured
in this case and the E2 cross section is comparable in magnitude with
the M1 cross section.

The interplay between longitudinal and transverse E2 components in
the forward direction can be seen in more details in Fig. \ref{fig3},
where the PWBA E2 cross sections  are plotted versus the incident
electron energy for different small scattering angles. The PWBA cross
sections are obtained from the same realistic RPA transition
densities of the considered 1$^+$ excitation in $^{156}$Gd, which
were used to obtain the DWBA results presented in Figs. \ref{fig1}
and \ref{fig2}. It is seen from Fig. \ref{fig3} that the transverse
E2 cross section (dotted lines) is almost constant and independent of
the incident energy $E_i$, while the Coulomb cross section (dashed
curves) is increasing quadratically with $E_i$.

The E2 cross section is purely transverse for $\theta = 0.1^{\circ}$
and 0.5$^{\circ}$ in the whole range ($10 < E_i < 210$ MeV) of
incident energies studied. In contrast, the Coulomb term dominates
for $E_i > 150$ MeV at $\theta = 1^{\circ}$ and for $E_i > 30$ MeV at
$\theta = 5^{\circ}$. Its strong increase with $E_i$ in the latter
case is seen also in the DWBA E2 cross section plotted in Fig.
\ref{fig1}. This behaviour can be understood by rewriting the PWBA
cross section (\ref{pwba}) with the help of (\ref{qe}) and
(\ref{vltq}): 
\begin{eqnarray}
\biggl ( {d \sigma \over d \Omega} \biggr )_{\rm PWBA} = \biggl (
{\alpha Z \over q} \biggr ) ^2 \Bigl \{ W_{\ell} \vert F^C_L (q) 
\vert ^2 
+ W_t \Bigl [ \vert F^E_L (q) \vert ^2  + \vert F^M_L (q) \vert ^2 
\Bigr ] \Bigr \},  \nonumber \\ 
 W_{\ell} = \cot ^2 (\theta /2), \quad 
 W_t = { 1 +  \sin ^2 (\theta /2) \over 2 \sin ^2 (\theta /2)}, 
 \quad W_t/W_{\ell} = V_T.
 \label{pwbq}
\end{eqnarray}

The $q$-dependence of the form factors $\vert F_L(q) \vert ^2$
(\ref{pwbq}) is estimated in PWBA by taking the leading terms (lowest
$L$) of the corresponding multipole operators into account
\cite{willey}. The Bessel functions are approximated afterwards 
in the long-wave limit (small $q$) as $j_L(qr) \sim (qr)^L$, so that
their contribution to the form factors is $\sim (qR)^L$
\cite{bar63,willey,uber}, where $R$ is the nuclear radius. The
transverse operators have convection and spin parts (\ref{jden}). It
turns out that both parts exhibit the same leading $q$-dependence for
magnetic, but not for electric operators \cite{uber}.

The choice of Willey \cite{willey} for the leading electric term is 
objected by \"Uberall \cite{uber} who notes that only the spin term
was considered in \cite{willey}. Willey has, however, both terms in
view when deciding to neglect $E_x /q \hbar c$ in favour of $q \hbar
c /M c^2$, where $M$ is the nucleon mass. Thus, the choice of Willey
will be correct for $q \gg q_0$, where 
\begin{equation}
q_0^2 = E_x Mc^2 /(\hbar c)^2 = 0.024 E_x \mbox{ fm}^{-2}
\label{qnot}
\end{equation}
and the excitation energy $E_x$ must be supplied in MeV. We obtain
$q_0 = 0.27$  fm$^{-1}$ for the low-energy orbital M1 excitations with
$E_x \approx 3$ MeV, studied here. Therefore, in our case we agree
with the choice of \"Uberall for the leading convection electric
term in the low-$q$ limit, while the choice of Willey remains valid
for the spin current where the problem of two competing leading terms
does not arise.

After introducing the above discussed correction, we obtain in the
low-$q$ limit from the expressions (2.64) for multipole operators in
Ref. \cite{willey} the following estimates for the $q$-dependence of
the PWBA form factors (\ref{pwbq}):
\begin{eqnarray}
\vert F^C_L (q) \vert ^2 \sim (qR)^{2L}, \quad 
\vert F^E_L (q) \vert ^2 _{\rm conv} \sim \biggl ( {E_x \over q 
\hbar c} \biggr ) ^2 (qR)^{2L}, \nonumber \\
\vert F^E _L (q) \vert ^2 _{\rm spin} \sim \biggl ( {q \hbar c \over 
Mc^2 } \biggr ) ^2 (qR)^{2L}, \quad 
\vert F^M_{L-1} (q) \vert ^2 \sim \biggl ( {q \hbar c \over 
Mc^2 } \biggr ) ^2 (qR)^{2L-4}.  
\label{fflow}
\end{eqnarray}
Upon insertion of (\ref{fflow}) into (\ref{pwbq}), we get the
following estimates for the PWBA cross sections in the low-$q$ limit:
\begin{equation}
   \begin{array}{l l l}
d \sigma (CL) & \sim  (qR)^{2L-2} R^2 W_{\ell}, & \\ 
d \sigma (EL,{\rm conv}) & \sim  \biggl ( \displaystyle\frac{E_x} 
{\hbar c}  \biggr )^2 (qR)^{2L-4} R^4 W_t, \; \; & \mbox{for } q < q_0, 
 \\ 
& \sim  d \sigma (EL,{\rm spin}), \; \; & \mbox{for } q > q_0, 
 \\ 
d \sigma (EL,{\rm spin}) & \sim  \biggl ( \displaystyle\frac{\hbar c} 
{M c^2} \biggr )^2 (qR)^{2L} W_t, &  \\
d \sigma (ML) & \sim  \biggl ( \displaystyle\frac{\hbar c} {M c^2} 
\biggr )^2 (qR)^{2L-2} W_t, & 
  \end{array} \label{sigl} 
\end{equation}
where $q_0$ is defined in (\ref{qnot}). The expression for the
magnetic cross section applies to both convection and spin current
contributions.  Although the above estimates (\ref{sigl}) are in
agreement with previous results \cite{willey,uber}, the
$q$-dependence in (\ref{sigl}) is qualitatively different. This can
be seen by comparison, e.g. with Table XXIV of \cite{uber}. The
difference arises from the $q$-dependence of $E_i^{-2}$ (\ref{pwba}),
expressed explicitly in (\ref{pwbq}) through (\ref{qe}). The previous
PWBA estimates for the $q$-dependence of the cross sections are valid
only for a fixed incident energy, i.e. they are meaningful when the
angular dependence of the form factor is studied for a constant
incident energy. The scattering angle is however fixed in modern
experiments \cite{achimrev,ramfagg} and the form factor is studied by
measurements for different incident energies. One has to apply in
this case the estimates (\ref{sigl}).  They are more general since
they are valid also for a fixed incident energy. The angular
dependence is given in this case by the kinematic factors $W$
(\ref{pwbq}).

We obtain from (\ref{sigl}) the following estimates for the
considered M1 and E2 PWBA cross sections by choosing the low-$q$
limit of the convection E2 excitation:
\begin{eqnarray}
d \sigma (C2) \sim q^2 R^4 W_{\ell}, \quad 
d \sigma (E2,{\rm conv}) \sim \biggl ( {E_x \over \hbar c} \biggr )^2 
R^4 W_t, \nonumber \\
d \sigma (E2,{\rm spin}) \sim \biggl ( {\hbar c \over M c^2} \biggr )^2 
(qR)^4 W_t , \quad 
d \sigma (M1) \sim \biggl ( {\hbar c \over M c^2} \biggr )^2 W_t.
\label{sig21}
\end{eqnarray}
Turning back to the inspection of Fig. \ref{fig3}, we see from
(\ref{sig21}) that the quadratic $q$-dependence of the Coulomb E2
excitation (dashed curves) agrees with the PWBA prediction
(\ref{sig21}). Moreover, it becomes clear that the constant
transverse E2 cross section (dotted lines) must originate from a
dominant convection current, because the transverse spin current
would exhibit a strong $q^4$-dependence. Let us note that the
previous estimates \cite{willey,uber} would not be able to explain
the constant transverse cross section of Fig. \ref{fig3}, because
they predict $q^2$ and $q^6$ dependencies for the convection and spin
E2 cross sections, respectively.  

The following qualitative estimates for the relative transverse 
contributions are obtained directly from (\ref{sigl}):
\begin{eqnarray}
{d \sigma (EL,{\rm conv}) \over  d \sigma (EL,{\rm spin}) } \approx
\biggl ( { E_x Mc^2 \over (q \hbar c)^2 } \biggr )^2 > 1,  \quad
\mbox{for } q < q_0,  \label{see}  \\ 
{ d \sigma (M,L-1) \over d \sigma (EL,{\rm spin}) }  \approx 
(qR)^{-4}, \label{smespi} \\ 
{ d \sigma (M,L-1) \over d \sigma (EL,{\rm conv}) } \approx 
\biggl ( {(\hbar c)^2 \over E_x R^2 Mc^2 } \biggr )^2 \approx
E_x^{-2}, 
\label{sme}
\end{eqnarray}
where $q_0$ is defined in (\ref{qnot}) and a nuclear radius $R=6.5$
fm was assumed for the rare-earth region to evaluate numerically the
latter relationship in (\ref{sme}), valid when the excitation energy
$E_x$ is given in MeV.  The ratios (\ref{see})-(\ref{sme}) should
hold for any scattering angle. They apply to both the convection and
spin parts of the magnetic excitation $(M,L-1)$.  It is seen from
(\ref{see}) that the convection electric cross section dominates over
the spin electric one since it is taken from (\ref{sigl}) just in the
low-$q$ limit $q < q_0$, necessary to ensure its dominance.  Hence,
the longitudinal cross section has to be compared only with the
transverse magnetic and convection electric ones:
\begin{eqnarray}
{ d \sigma (CL) \over d \sigma (EL,{\rm conv}) } \approx 
\biggl ( {q \hbar c \over E_x } \biggr )^2 { 1 \over V_T},  
 \label{sce} \\ 
{ d \sigma (CL) \over d \sigma (M,L-1) } \approx \biggl ( {R Mc^2 
\over \hbar c } \biggr )^2 { (qR)^2 \over V_T}. 
\label{scm}
\end{eqnarray}
After inserting $V_T(0^{\circ})$ from (\ref{vtzero}) in the above
relationships and taking into account that $q \hbar c = E_x$ at the
photon point, one can easily verify that the cross section is
dominated by transverse excitations at $\theta = 0^{\circ}$ and the
longitudinal contribution is negligible:
\begin{eqnarray}
{ d \sigma (CL) \over d \sigma (EL,{\rm conv}) } (0^{\circ}) \approx 
2 \biggl ( {m_e c^2 \over E_i } \biggr )^2 \ll 1, \nonumber \\ 
{ d \sigma (CL) \over d \sigma (M,L-1) }  (0^{\circ}) \approx 
2 \biggl ( {M m_e c^4 R^2 E_x \over (\hbar c)^2 E_i } \biggr )^2 \approx 
2 \biggl ( {m_e c^2 E_x \over E_i } \biggr )^2 \ll 1,
\label{scmzero}
\end{eqnarray}
where the same numerical estimate was used in the last relationship
as in (\ref{sme}). This particular case of extreme forward scattering
was discussed already above. Let us check now the predictions
(\ref{smespi})-(\ref{scm}) by comparison with realistic results. 

The PWBA cross sections  displayed in Fig. \ref{fig4} are plotted
versus the transferred momentum $q$ for $\theta = 70.53^{\circ}$.
This angle ensures $V_T = 1$, as seen from Eq. (\ref{vlt}), i.e. a
kinematic condition which is equally favourable for both longitudinal
and transverse excitations. The PWBA plots represent therefore the
form factors $\vert F_L(q) \vert ^2$ (\ref{pwba}), (\ref{pwbq}) up to
a coefficient common for all of them. The cross sections are
calculated with our microscopic RPA transition densities for the
strongest low-lying orbital 1$^+$ excitation in $^{156}$Gd. The ratio
between the orbital and spin transition matrix elements of the M1
operator \cite{nofa88} is $R_{o.s.}$ = 7.8 for this excitation, i.e.
it has a predominant orbital nature.

The PWBA M1 cross section is displayed as a continuous curve in the
top plot of Fig. \ref{fig4}, together with the orbital contribution
alone (dashed curve). This corresponds to the decomposition of the
current transition density ${\cal J}^{\nu} _{11} (r)$  (\ref{jden})
into convection (orbital) and magnetization (spin) parts. It is seen
that the M1 cross section is almost purely orbital in the considered
$q$-range with negligible spin contributions. This is a typical
property of the studied orbital 1$^+$ excitations in heavy deformed
nuclei \cite{noj95}. The predominant orbital nature at the photon
point (see $R_{o.s.}$ above) is preserved up to $q = 1.4$ fm$^{-1}$,
as seen from Fig. \ref{fig4}. 

This typical feature, determined by the nuclear dynamics, is not
present in the qualitative PWBA estimates (\ref{fflow}),
(\ref{sigl}), where the nuclear orbital, spin, and charge transition
matrix elements are assumed to be all of comparable magnitude.
The transverse E2 cross section should be one order of magnitude
larger than the M1 cross section, as predicted by (\ref{sme}) for the
considered 1$^+$ excitations with $E_x \approx 3$ MeV. Comparison of
the upper two plots in Fig. \ref{fig4} invalidates this PWBA
prediction because the M1 cross section is two orders of magnitude
larger than the transverse E2 contribution. This is observed also in
the plots of Fig. \ref{fig5}, where the M1 cross section is larger
than the E2 contribution for small scattering angles. The E2 cross
section arises in this case mainly from the transverse E2 excitation,
as one can check from Fig. \ref{fig3} for small angles and small
incident energies.

Let us verify now the predictions (\ref{sce}), (\ref{scm}). It is seen
from the middle plot of Fig. \ref{fig4} that for small $q$ the PWBA
Coulomb (C2) cross section (continuous curve) is more than one order
of magnitude larger than the transverse E2 cross section (dashed
curve), as it should be expected from (\ref{sce}) for $E_x \approx 3$
MeV.

Comparison of the upper two plots in Fig. \ref{fig4} shows that the
M1 response dominates over the C2 cross section only for small
momentum transfer $q < 0.17$ fm$^{-1}$ ($E_i < 30$ MeV). This
crossing point, where the M1 and C2 cross sections are equal, is
confirmed also by more realistic DWBA results, displayed in the
bottom plot of Fig. \ref{fig4} (note that they are plotted versus
$q_{\rm eff}$). On the other hand, Eq. (\ref{scm}) predicts that the
crossing should take place for $q = 0.005$ fm$^{-1}$ or $E_i = 0.85$
MeV, which is impossible since the incident energy must be larger
than the excitation energy $E_x \approx 3$ MeV. Therefore, the PWBA
prediction (\ref{scm}) overestimates the C2 with respect to the M1
cross section by three orders of magnitude. 

In order to study this problem in more detail, we compare in Fig.
\ref{fig5} M1 and E2 DWBA cross sections plotted versus the
scattering angle. They are sections of the two plots in Fig. \ref{fig1} 
for different fixed values of the incident electron energy $E_i$. It
is seen from Fig. \ref{fig5} that the M1 cross section (continuous
curves) is larger than the E2 cross section (dashed curves) in the
whole kinematical region for small incident energies $E_i < 30$ MeV.
Above this energy the E2 response is dominant for almost any angle,
except for very small (forward) or very large (backward) angles. One
can check that the crossing point of the two curves at small angles
is characterized by $(qR)^2 \approx V_T$ for different incident
energies. Thus, the qualitative estimate (\ref{scm}) is not confirmed
and the M1 response is dominant up to a much higher momentum transfer
than predicted by PWBA.

Small scattering angles are not favourable for the study of nuclear
excitations through inelastic electron scattering because of the
large background originating from the radiative tail of the strong
elastic peak \cite{willey,uber,fagg}. On the other hand, a technique
was developed for precise measurement of the inelastic cross section
relative to the elastic peak, which can not be applied for backward
angles \cite{fagg}. There are many other sources of radiative
background, contributing at 180$^{\circ}$ as well. Among them, the
magnetic bremsstrahlung is particularily strong \cite{fagg}.
We turn our attention to the backward direction where most of the
experiments on magnetic excitations have been done.

Let us check the estimates (\ref{smespi})-(\ref{scm}) in the case of
backward scattering where the dominant longitudinal contribution is
strongly damped by the kinematics. The expression (\ref{vlt}) for the
kinematic factor $V_T$ diverges at 180$^{\circ}$. Its correct
limiting behaviour \cite{theis} can be obtained from Eqs. (4-12b,c)
of \cite{uber} by taking the electron rest mass in the expression for
$V_{\ell}$ into account. The transverse factor $V_t$ (\ref{vlt}) does
not need such a correction at backward angles, so that,
\begin{eqnarray}
V_{\ell}^e = { \cos ^2 (\theta /2) + (m_ec^2)^2 /(E_iE_f) \over 4 \sin
^4 (\theta /2) }, \quad 2 V_T^e = { 1 + \sin ^2 (\theta /2) \over 
\cos ^2 (\theta /2) + (m_ec^2)^2 /(E_iE_f) },  \label{vlpi} \\
V_T^e (180^{\circ}) = \biggl ( {E_i \over m_e c^2 } \biggr )^2
  = 2 V_T (0^{\circ}),  
\label{vtpi} 
\end{eqnarray}
where $V_T (0^{\circ})$ is given by (\ref{vtzero}).  The upper index
$"e"$ is to remind that the electron mass was taken into account.
The corrected factors (\ref{vlpi}) differ from the rough estimate
(\ref{vlt}) only for small incident energies and angles close to
180$^{\circ}$. 
One obtains from (\ref{sce}), (\ref{scm}) and (\ref{vtpi}) for a full
backward angle:
\begin{eqnarray}
{ d \sigma (CL) \over d \sigma (EL,{\rm conv}) } (180^{\circ}) 
\approx \biggl ( { 2 m_ec^2 \over E_x } \biggr )^2 ,  \nonumber \\ 
{ d \sigma (CL) \over d \sigma (M,L-1) }  (180^{\circ}) \approx 
\Bigl (2M m_e c^4 \Bigr )^2 \biggl ( { R \over \hbar c } \biggr )^4
\approx 1.1. 
 \label{scepi} 
\end{eqnarray}
It is seen from (\ref{scepi}) that the C2 and M1 cross sections
should be of comparable magnitude even in fully backward scattering.
The dominant contribution arises however from the transverse
convection E2 response. According to (\ref{scepi}), this contribution
should be one order of magnitude larger than the longitudinal C2
cross section for the considered excitation energies $E_x \approx 3$
MeV.

The above qualitative PWBA estimates have been made without disposing
of any information about the dynamics of the considered nuclear
excitations. Let us compare them with our PWBA results obtained from
the realistic RPA transition densities of the studied orbital 1$^+$
excitations in heavy deformed nuclei. The PWBA cross sections are
plotted in Fig. \ref{fig6} for $\theta = 179.5^{\circ}$. The M1
cross section arising from the convection current alone (dashed 
curve) is shown in the upper plot together with the total (convection
plus spin, continuous curve) M1 cross section. It is seen that the
M1 response is almost purely orbital for small momentum transfer, but
the spin M1 contributions become larger for $q > 1.4$ fm$^{-1}$. 

The PWBA E2 cross sections, displayed in the lower plot of Fig.
\ref{fig6}, demonstrate that the C2 response is already negligible
at this large angle. The transverse E2 cross section is almost two
orders of magnitude smaller than the M1 cross section for small
momentum transfer $q$. These results, obtained from a realistic RPA
wave function, stand in contrast with the general PWBA expectations
(\ref{scepi}) where the nuclear dynamics is not taken into account.
According to (\ref{scepi}), the transverse E2 excitation arising from
the convection current should dominate over the M1 and C2 responses
for small $q$ in backward scattering.

The displayed cross sections are calculated with the simple
kinematic factors (\ref{vlt}) but the above conclusions are not
altered when the more precise longitudinal factor $V_{\ell}^e$
(\ref{vlpi}) is used instead of $V_{\ell}$ to calculate the C2 cross
section from the bottom plot of Fig. \ref{fig6}.  The two expressions
differ considerably only for small incident energies and angles very
close to 180$^{\circ}$.  But even for $\theta = 179.5^{\circ}$ and
realistic incident energies $E_i > 20$ MeV the largest discrepancy
(for $E_i = 20$ MeV) is not very relevant. The correct longitudinal
factor $V_{\ell}^e$ (\ref{vlpi}) is about 40 times larger than the
rough value (\ref{vlt}), but $V_{\ell}^e$ is still 1270 times smaller
than the transverse factor $V_t$. Thus, also with the correct factor
$V_{\ell}^e$ the situation characterized by a negligible longitudinal
response is not altered: the C2 contribution is still one order
of magnitude smaller than the E2 cross section for small $q$ ($E_i <
10$ MeV). 

\section{PWBA analysis for large momentum transfer}

The intensity of the scattered electrons decreases several
orders of magnitude in the backward direction but this kinematical
region is more interesting because the maximal momentum  is
transferred there for a given incident energy, as seen from
(\ref{qe}). One can investigate in this way the high-$q$ region, far
from the photon point which could be studied also with more precise
methods as photonuclear reactions \cite{kneisrev}.

The contribution of the Bessel functions to the form factors
(\ref{pwba}) are approximated in the high-$q$ limit  \cite{uber} by
$j_L(qR) \sim 1$.  This corresponds to $q > 0.15$ fm$^{-1}$ for the
considered rare-earth nuclei. Let us assume additionally that $q >
0.27$  fm$^{-1}$, see Eq. (\ref{qnot}), so that the leading
convection and spin terms in the transverse E$L$ and M$L$--1 form
factors have the same $q$-dependence. The estimates (\ref{fflow})
acquire now the simpler form:
\begin{eqnarray}
\vert F^C_L (q) \vert ^2 \sim 1, \quad \nonumber \\ 
\vert F^E _L (q) \vert ^2 \sim \vert F^M_{L-1} (q) 
\vert ^2 \sim  \biggl ( {q \hbar c \over Mc^2 } \biggr ) ^2 . 
\label{fflar}
\end{eqnarray}
One obtains from (\ref{fflar}) and (\ref{pwbq}) the following
estimates for the PWBA cross sections for large momentum transfer:
\begin{eqnarray}
d \sigma (CL)  \sim  q^{-2} W_{\ell}, \quad 
d \sigma (EL) \sim d \sigma (M,L-1) \sim  \biggl ( \frac{\hbar c} 
{M c^2} \biggr )^2 W_t, \label{siglar} \\ 
{ d \sigma (EL) \over d \sigma (M,L-1) } \approx 1, 
  \label{emla} \\ 
{ d \sigma (CL) \over d \sigma (EL) } \approx 
{ d \sigma (CL) \over d \sigma (M,L-1) } \approx 
\biggl ( {Mc^2 \over q \hbar c } \biggr )^2 { 1 \over V_T}, 
\label{sclar} \\
{ d \sigma (CL) \over d \sigma (EL) } (180^{\circ})  \approx 
{ d \sigma (CL) \over d \sigma (M,L-1) } (180^{\circ})  \approx 
\biggl ( {M m_e c^4 \over 2 E_i ^2 } \biggr )^2 , 
\label{sclpi}
\end{eqnarray}
where (\ref{vtpi}) and (\ref{qeff}) were used to get (\ref{sclpi}).
If we choose an angle where $V_T =1$, the ratio (\ref{sclar}) is
always a large number since $q < Mc^2 /(\hbar c)$ = 4.75 fm$^{-1}$.
This is indeed the condition for validity of the non-relativistic
treatment of the nucleus \cite{willey}. One can conclude in this way
from the estimates (\ref{emla}), (\ref{sclar}), that the transverse
E2 and M1 cross sections should be of comparable magnitude, while the
longitudinal C2 excitation should dominate over them for not very
large backward angles. 

The prediction (\ref{sclar}) is confirmed in Fig. \ref{fig5} where
the E2 response (mainly longitudinal C2) dominates over M1 for
large $q$ ($E_i > 30$ MeV) apart from very large backward angles.
This qualitative conclusion is confirmed also in Fig. \ref{fig4},
where the relative magnitudes of the form factors are clearly seen
due to $V_T = 1$. The longitudinal C2 dominates over the transverse
E2 and M1 responses in Fig. \ref{fig4} for large transferred
momentum. Let us remind that the C2 dominance over the transverse E2
excitation, predicted by (\ref{sce}) for low $q$, was confirmed in
the previous section, i.e. this relationship is valid for a wide
region of transferred momenta.

The longitudinal dominance over the transverse currents is usually
considered as a signature for collective excitations
\cite{bar62,heis}. One should note, however, that this dominance is
predicted by Eqs. (\ref{sce}), (\ref{scm}) and (\ref{sclar}), which
are not restricted by any assumptions about the nature or extent of
collectivity of the nuclear transition. On the other hand, collective
excitations have been associated usually with the assumption for an
irrotational, incompressible flow \cite{heis81}, whose signature is
the vanishing transverse electric current ${\hat {\cal J}} _{L,L+1,M}
(r)$ (\ref{mdop}). Our RPA transition densities for the transverse E2
currents with $L^{\prime}=L-1$ and $L^{\prime}=L+1$ are of
comparable magnitude (not in favour of collectivity according to the
latter criterium) but their amplitude is one order of magnitude
smaller than the amplitude of the C2 transition density (in favour of
collectivity according to the former criterium). On the basis of
different considerations we came to the conclusion
\cite{nofa88,noj95} that the low-lying orbital 1$^+$ excitations
represent a weakly collective scissor mode.

In contrast to the validity of (\ref{sclar}), the relationship
(\ref{emla}), which should hold for any angle, is not confirmed by
realistic results.  Decomposition is made in  Figs. \ref{fig4} and
\ref{fig6} for 70.53$^{\circ}$ and 179.5$^{\circ}$, respectively. It
is seen from these two figures that for large $q$ the transverse E2
excitation (mainly convection) is almost two orders of magnitude
smaller than the M1 response (also mainly convection). It was found in
the previous section that the corresponding prediction (\ref{sme})
for this ratio at low $q$ contradicts our realistic results in the
same way, but the discrepancy is three orders of magnitude for small
momentum transfer.

The condition of M1 dominance over the C2 response for full backward
scattering and for large $q$ is obtained from (\ref{sclpi}): $E_i^2 >
Mm_e c^4/2$ or $E_i > 15.5$ MeV. This estimate is confirmed in Fig.
\ref{fig6}, where PWBA cross sections for $\theta = 179.5 ^{\circ}$
are displayed. They are obtained with realistic RPA transition
densities.  The lower plot shows that the transverse E2 contribution
dominates over the C2 cross section. It is seen from the comparison
of the two plots that also the M1 response dominates over the C2
excitation, as expected from (\ref{sclpi}). The upper plot
demonstrates that the M1 response is almost purely orbital for $q <
1.4$ fm$^{-1}$: it originates mainly from the convection current
(dashed curve).  The spin M1 current becomes more important above
this momentum transfer.  This is due to the fact that the convection
transition density is peaked close to the nuclear surface (at $r = 5$
fm) and almost vanishes for $r < 2$ fm, while the spin transition
density has an appreciable amplitude also deep inside the nucleus, a
region reached at high momentum transfer.

It has been argued \cite{fagg,ramfagg} that the spin-flip M1
response should dominate in backward scattering when the electron
rest mass can be neglected in comparison with its incident energy and
the transferred momentum is much larger than the nuclear excitation
energy.  These approximations are reasonable in backward scattering
for not very small incident energy. In this case the electron can be
viewed as a massless particle with a good helicity \cite{sitenko},
i.e. it is longitudinally polarized and the electron spin is aligned
along or opposite to its momentum.  The scattering to 180$^{\circ}$
is considered in Ref. \cite{fagg} as an occurrence of spin-flip, so
that the longitudinal and transverse convection parts of the
interaction should be strongly damped and only transverse spin M1 and
spin E2 interactions will compete.

The ratio of the corresponding cross sections is given by the PWBA
estimates (\ref{smespi}) for small $q$ and (\ref{emla}) for large
$q$. Hence, if only transverse spin interactions had to compete with
each other, a dominant spin M1 response could be expected from the
above helicity arguments only for small $q < 0.15$ fm$^{-1}$  where
the necessary high-momentum approximation is not justified very well.
Moreover, so small $q$-values are not reached in the $(e,e^{\prime})$
experiments \cite{bohle84,achimrev} investigating M1 excitations in
heavy deformed nuclei, where $E_i \ge 20$ MeV or $q \ge 0.2$ fm$^{-1}$
in backward scattering. The spin M1 and E2 cross sections should be
almost equal at high $q$ according to (\ref{emla}).

Inspection of Fig. \ref{fig6} demonstrates that in backward
scattering (negligible longitudinal response) the convection M1
cross section is much larger than the transverse E2 contribution even
for large momentum transfer up to $q = 1.2$ fm$^{-1}$.  In this way,
both the qualitative PWBA prediction (\ref{emla}) and the above
helicity arguments for a dominant spin M1 response contradict the
realistic results for the considered orbital M1 transition.  The
equality of the transverse M1 and E2 cross sections is predicted in
(\ref{emla}) to take place for any $q > 0.15$ fm$^{-1}$ while it
is reached in reality for a much higher momentum transfer $q = 1.2$
fm$^{-1}$. It is true that the longitudinal interaction is strongly
suppressed in backward scattering. This is easily seen from the small
longitudinal kinematic factor $V_{\ell}^e$ (\ref{vlpi}). It is also
true that high-energy electrons are longitudinally polarized. But
even in the PWBA treatment of backward scattering of longitudinally
polarized electrons both convection and spin currents contribute on
equal footing to the cross section  \cite{rose}.

Let us compare finally the M1 and E2 DWBA cross sections plotted in
Fig. \ref{fig7} versus the effective momentum transfer $q_{\rm eff}$ 
(\ref{qeff}) for different scattering angles. The M1 cross section is
larger than the E2 contribution in the whole region of incident
energies studied ($E_i < 210$ MeV) for small forward angles
$\theta \le 5 ^{\circ}$ (not shown in the figure, but seen in Fig.
\ref{fig1}). At larger scattering angles there is always a crossing
point for a given momentum transfer, beyond which the E2 response is
dominant. The $q$-value of the crossing point obeys the relationship
discussed in the previous section.

It is seen from Fig. \ref{fig7} that the large $q$-values lie beyond
the crossing point where the E2 excitation (mainly longitudinal) is
dominant over M1. However, the crossing point moves towards higher
$q$-values when approaching the backward direction, so that the M1
response becomes more important even for moderately large
momentum transfer. The two cross sections decouple only at
175$^{\circ}$. At this angle the M1 excitation dominates up to
$q_{\rm eff} = 1.3$ fm$^{-1}$, while the C2 and the transverse E2
responses are of comparable magnitude. The C2 contributions are
negligible beyond 175$^{\circ}$ and the total cross section is
determined only by the relative transverse M1 and E2 responses which
are multiplied by the same transverse factor $V_t$ (\ref{vlt}).

Cross sections of lower multipolarity decrease faster for larger $q$
\cite{bar62}.  According to (\ref{sigl}), the cross sections obey the
following order of increasing multipolarity in their momentum
dependence: $d \sigma (M,L-1) \sim (qR)^{2(L-2)}, \; d \sigma (CL)
\sim (qR)^{2(L-1)}, \; d \sigma (EL) \sim (qR)^{2L}$, where the
high-$q$ limit, common for convection and spin transverse electric
responses, is chosen.  In addition to this general trend, our RPA
results show that the studied orbital excitations have typical
convection M1 transition densities similar to the charge C2
transition density: they are characterized by a well-localized single
bump. In contrast, the spin M1 and transverse E2 transition densities
exhibit more oscillations. Hence, the spin M1 response decreases less
for larger $q$ in comparison with the convection M1 term, i.e. the
spin response behaves effectively as a higher multipolarity with
respect to $q$.

One should expect, therefore, that the transverse E2 response will
dominate over the transverse M1 excitation for very large transferred
momenta.  This can be seen in the two bottom plots of Fig. \ref{fig7}
for $q_{\rm eff} > 2$ fm$^{-1}$. Thus, the transverse E2 response
will be dominant even in a fully backward scattering for very large
momentum transfer.

\section{Summary} 

We have studied the $(e,e^{\prime})$ cross sections of low-lying
excitations with $K^{\pi} = 1^+$ in heavy deformed nuclei. They have
a predominantly orbital nature, in contrast to the more extensively
explored spin-flip M1 excitations in spherical nuclei. Moreover, we
have found recently \cite{nfd94,dnf95} that the accompanying E2
transitions with $I^{\pi}K = 2^+1$ provide appreciable contributions
to the measured M1 cross sections in heavy deformed nuclei even in
backward scattering where the M1 response is expected to dominate.

We obtain the cross sections from realistic microscopic RPA
transition densities.  The low-lying 1$^+$ excitation with the
strongest orbital M1 transition in $^{156}$Gd is studied as a typical
example for such excitations which have been interpreted as a weakly
collective scissors mode \cite{nofa88,noj95}.  The already reported
good agreement \cite{nfd94,dnf95} of the theoretical DWBA cross
sections with the available experimental data ($\theta = 165
^{\circ}$) confirms the importance of the accompanying E2 transitions
in heavy deformed nuclei and places some confidence in the validity
of the theoretical cross sections for different scattering
conditions.

The dependence of the theoretical M1 and E2 DWBA cross sections upon 
the scattering angle $\theta$ and the incident electron energy $E_i$
is studied here in the whole kinematical region $0 ^{\circ} < \theta
< 180 ^{\circ}$ for $10 \le E_i \le 210$ MeV. In some special cases
the cross sections are calculated also within PWBA in order to
separate longitudinal from transverse contributions (not possible in
DWBA).  The DWBA and PWBA cross sections, obtained from realistic RPA
transition densities, are compared to qualitative PWBA predictions
which ignore completely the nuclear dynamics. They are based on
general considerations of the scattering kinematics and the momentum
dependence of the unknown form factors. The latter are fully
specified by our RPA transition densities which is the main advantage
of the microscopic approach.

The theoretical cross sections exhibit the following peculiarities in
agreement with the qualitative PWBA predictions: 

  i) For a given momentum transfer and $\theta > 90 ^{\circ}$ the
(purely transverse) M1 cross section is almost constant with
$\theta$, while the E2 cross section decreases fast in the backward
direction. This effect is simply due to the interplay between the
longitudinal and transverse kinematic factors.

  ii) For small momentum transfer $q$ and a fixed scattering angle
$\theta$ the transverse convection M1 and E2 cross sections are
almost independent of $q$, while the longitudinal C2 cross section
increases as $q^2$. The convection current is dominant in the
transverse E2 response for small $q$. 

  iii) The longitudinal C2 (Coulomb) form factor is much larger than
the transverse M1 and E2 form factors. Therefore, the C2 response 
is dominant if the longitudinal kinematic factor is not very small.
This condition is met away from the extreme forward and backward
directions. Even for very small angles  the Coulomb suppresion is no
more effective at high incident energies because of the momentum
dependence in the kinematic factor.

  iv) For scattering angles close to 180$^{\circ}$, where the Coulomb
response is negligible, the transverse E2 excitation dominates 
over the M1 transition at high transferred momenta. This is due to
the fact that cross sections of lower multipolarity decrease faster
with the increase in $q$.

The longitudinal dominance  over the transverse currents is usually
considered as a signature for collective excitations. However, this
dominance is predicted in PWBA by Eqs. (\ref{sce}), (\ref{scm}) and
(\ref{sclar}), whose derivation does not involve any assumptions
about the nature or extent of collectivity of the nuclear transition.

The PWBA predictions are not able to provide reliable numerical
estimates for the limiting values of angles, incident energies and
transferred momenta which specify the validity range of different
predicted relationships. The qualitative PWBA estimates are either
not able to predict or contradict the following realistic microscopic
results for low-lying orbital 1$^+$ excitations:

   v) The ratio between the C2 and M1 cross sections is three orders
of magnitude smaller than estimated in PWBA. The PWBA predictions for
a strong Coulomb response could be more appropriate for rotational
transitions but not for the orbital M1 excitations considered here.
Thus, a more pronounced M1 dominance is present in our results at
small momentum transfer, especially in backward scattering. The M1
response is dominant at any angle for incident energies $E_i < 30$
MeV. It is dominant in the forward direction, e.g. for $E_i < 150$ at
$\theta = 1^{\circ}$ and even up to larger energies at smaller
angles.
   
   vi)  The longitudinal C2 excitation is comparable with the
transverse E2 response at $\theta = 175 ^{\circ}$ but becomes
negligible beyond this angle in the backward direction. For such
large angles the total cross section is determined by the interplay
between transverse M1 and E2 excitations.

   vii) In backward scattering ($175 ^{\circ} < \theta < 180
^{\circ}$) the transverse M1 response is stronger than the transverse
E2 excitation up to $q = 1.2$ fm$^{-1}$. This feature stands in
contrast to the qualitative PWBA predictions for a dominant
transverse E2 transition in both low- and high-$q$ regions (the
latter refers to $q > 0.15$ fm$^{-1}$).  The two cross sections are
comparably large in the region $1.2 < q < 2$ fm$^{-1}$ and the
transverse E2 cross section is expected to dominate for $q > 2$
fm$^{-1}$ ($E_i > 200$ MeV in backward scattering).

   viii) The M1 response originates mainly from the convection
current up to $q = 1.4$ fm$^{-1}$.  The spin M1 current becomes more
important above this momentum transfer due to the volume character of
the oscillating spin M1 transition density.

The above properties apply to the low-lying 1$^+$ excitations in
heavy deformed nuclei studied here. They are mainly orbital with
small spin contributions. Such typical properties can not be
predicted without information about the wave function of the
considered nuclear excitation. Though high-energy electrons can be
viewed as longitudinally polarized massless particles with a good
helicity, there is no suppression of the convection current
interaction in backward scattering. This is confirmed by the studied
orbital excitations where the convection current provides the main
contribution to the M1 response up to a high momentum transfer.

\bigskip 
 Thanks are due to Henk P. Blok and Jochen Heisenberg for providing
us with their DWBA code and useful communications on the underlying
fromalism. This work is supported by the Deutsche
Forschungsgemeinschaft (DFG) under contracts Fa 67/15-1 and Fa
67/15-2.

  \newpage
\centerline{\bf Appendix}

The transition densities (\ref{rhoden}), (\ref{jden}) are
calculated with the RPA wave functions in the laboratory frame in the
way described in \cite{fanota} for M1 transitions. In the case of E2
transitions they have the form:
\begin{eqnarray}
\rho _2 ^{\nu} (r) = {1 \over 2} \sum _{0 < i < f} 
  \Bigl [ F^{\nu} _{+1} (fi) \rho _{21} (fi,r) +  F^{\nu} _{+1} 
(f\tilde{i}) \rho _{21} (f\tilde{i},r) \Bigr ], 
\label{rhod} \\  
{\cal J} _{2L^{\prime}} ^{\nu} (r) = {1 \over 2} \sum _{0 < i < f} 
 \Bigl [ F^{\nu} _{-1} (fi) {\cal J} _{2L^{\prime}1} (fi,r)
+  F^{\nu} _{-1} (f\tilde{i}) {\cal J} _{2L^{\prime}1} 
(f\tilde{i},r) \Bigr ],  \label{jd} 
\end{eqnarray}
where the summation runs over single-particle states of the
Woods-Saxon potential with projections $K$ on the nuclear symmetry
axis $z$ obeying $0 < K_i < K_f$ and time-reversed states $\tilde{i}$
are treated explicitly. The nuclear dynamics is contained in the
factors $F^{\nu}_{\sigma}, \; \sigma = \pm 1$, which are linear
combinations of the RPA forward- and backward-going amplitudes $\psi
_{\nu} (fi,m)$ and $\phi _{\nu} (fi,m)$ \cite{nofa88}, respectively:
\begin{eqnarray}
 F^{\nu} _{\sigma} (fi) = \sqrt{2} \Bigl [  F^{\nu} _{\sigma} (fi,
 -1) -  F^{\nu} _{\sigma} (fi,  +1) \Bigr ], 
 \label{fup} \\
 F^{\nu} _{\sigma} (fi,m) = \psi _{\nu} (fi,m) + \sigma m \, 
 \phi _{\nu} (fi,m), 
\label{fsig}
\end{eqnarray}
The signature index $m = \pm 1$, corresponding to the
indistinguishable $x$ and $y$ directions, is of technical interest
only. It allows us to take symmetries into account and to avoid the
problem of symmetrizing the momentum operator \cite{willey}. Final
results do not depend on the signature, as seen from (\ref{rhod}), 
(\ref{jd}).  In order to save space we skip in (\ref{fup}),
(\ref{fsig}) and below expressions involving time-reversed initial
states $\tilde{i}$, though we always take such terms in the numerical
calculations into account.

The notations ${\rho}_{21} (fi,r)$ in (\ref{rhod}) and ${\cal J}
_{2L^{\prime}1} (fi,r)$ in (\ref{jd}) stand for the q.p. matrix
elements of the E2 charge and current density operators $\hat{\rho}
_{21} (r)$ and $\hat{\cal J} _{2L^{\prime}1} (r)$, respectively:
\begin{eqnarray}
{\hat \rho}_{21} (r) = \int Y_{21} (\Omega ) {\hat \rho} ({\bf r}) \;
d \Omega , \; \; 
{\hat {\cal J}} _{2L^{\prime}1} (r) = \int {\bf Y} _{2L^{\prime}1} 
(\Omega ) \bullet  {\hat {\bf J}} ({\bf r}) \; d \Omega,
\quad
L^{\prime} = 1,3,  \label{mdop}
\end{eqnarray}
${\bf Y} _{LL^{\prime}M} (\Omega )$ are the vector spherical
harmonics \cite{varsh}. The nuclear charge and current density
operators ${\hat \rho} ({\bf r})$ and ${\hat {\bf J}} ({\bf r})$ can
be found, e.g. in \cite{uber,fanota,willey}.

The q.p. matrix elements are obtained with the help of the
eigenfunctions $\Phi_i$ of the deformed  Woods-Saxon potential
\cite{dam69} in cylindrical coordinates ($\rho, z, \varphi$):
\begin{eqnarray}
\Phi_i ( K ^{\pi}, {\bf r} ) = \sum _j C^i_j \Psi_j (n_{\rho},
n_z,\Lambda,\Sigma = \textstyle{1 \over 2}, K, {\bf r})
\nonumber \\
 + \sum _{j^{\prime}} C^i_{j^{\prime}} \Psi_{j^{\prime}}
(n^{\prime} _{\rho}, n^{\prime}_z, \Lambda^{\prime} = \Lambda +1, 
\Sigma^{\prime} = - \textstyle{1 \over 2}, K, {\bf r}), \nonumber \\
\Psi (n_{\rho},n_z,\Lambda,\Sigma,{\bf r}) 
= \psi ^{\Lambda} (n_{\rho}, \rho ) \, \psi ( n_z , z) \, 
\psi (\Lambda , \varphi ) \, \chi (\Sigma ),   
 \label{wswf}
\end{eqnarray}
where $C^i_j, \; j = (n_{\rho}, n_z)$, are the coefficients of 
the expansion over the basis wave functions $\Psi_j$ of the 
axially symmetric harmonic oscillator. The cylindrical quantum
numbers $n_z, \, n_{\rho}$, correspond to the directions along and
perpendicular to the symmetry axis, while $\Lambda = K - 1/2$ and
$\Sigma$ are the projections of the orbital angular momentum and spin
operators on this axis. The angular integration over $d \Omega = \sin
\theta d\theta d \varphi$ in (\ref{rhod}), (\ref{jd}), reduces to 
integration over $-d \cos \theta$ since the integration over the 
azymuth angle $\varphi$ and the spin matrix elements can be
performed analytically in the basis (\ref{wswf}). One obtains in
this way the following expressions for the q.p. matrix elements of
the charge transition density (\ref{rhod}):
\begin{eqnarray}
\rho _{21} (fi,r) = e U_{+1} (f,i) \int Y_{21} (\Omega ) \Phi
^{\dag} _f ({\bf r}) \Phi _i ({\bf r}) d \Omega  \nonumber \\ 
= -e U_{+1} (f,i) \sqrt{ 15 \over 8 \pi} 
\sum _{\Lambda^{\prime} = \Lambda, \Lambda +1} 
\int {\rho z \over r^2} A^{fi} (\Lambda ^{\prime} +1, \Lambda 
^{\prime} ) \, d \> {\rm cos} \theta ,  \nonumber  \\
\rho = r \, {\rm sin} \theta , \quad z =  r \, {\rm cos} \theta , 
 \label{rhome}
\end{eqnarray}
where the coefficients $U_{+1}$ are linear combinations of the BCS
occupation numbers. The functions $A^{fi}$ contain only $\psi
^{\Lambda} (n_{\rho}, \rho ) \, \psi ( n_z , z)$ from (\ref{wswf}).
Thus, they depend only on $z$ and $\rho$, i.e. they are products of
Hermite and associate Laguerre polynomials \cite{nobofa}:
\begin{eqnarray}
U_{\sigma} (f,i) = u_f v_i + \sigma \, u_i v_f, 
\quad \sigma = \pm 1, \nonumber \\
A^{fi} (\Lambda_f, \Lambda_i ) = \sum_{kj} C^f_k (\Lambda_f) 
C^i_j (\Lambda_i) \psi ^{\Lambda_f}_k (\rho ) \, 
\psi ^{\Lambda_i}_j (\rho )  \> \psi_k (z) \psi_j (z).
\label{afi}
\end{eqnarray}

The q.p. matrix elements of the transverse E2 convection currents
have the form:
\begin{eqnarray}
{\cal J}^C _{211} (fi,r) = i \mu _N U_{-1} (f,i) {1 \over 2} 
\sqrt{3 \over \pi}  \sum _{\Lambda^{\prime} = \Lambda, \Lambda +1} 
\int \biggl \{ {z \over r} \Bigl [ {\partial \over \partial \rho_i }
A^{fi} (\Lambda ^{\prime} +1, \Lambda ^{\prime} ) \nonumber \\ 
- { \Lambda^{\prime} \over \rho } A^{fi} (\Lambda^{\prime} +1, 
\Lambda^{\prime} ) \Bigr ] 
+ {\rho \over r} {\partial \over \partial z_i } A^{fi} (\Lambda 
^{\prime} +1, \Lambda ^{\prime} ) \biggr \}  d \> {\rm cos} \theta , 
\label{jc21}
\end{eqnarray}
\begin{eqnarray}
{\cal J}^C _{231} (fi,r) = {i \mu _N \over \sqrt{2\pi}} 
U_{-1} (f,i) \sum _{\Lambda^{\prime} = \Lambda, \Lambda +1} 
\int \biggl \{ { (z^2 - 4 \rho ^2) \over r^3 } \Bigl [   
 z  {\partial \over \partial \rho_i } + \rho {\partial \over 
 \partial z_i } \Bigr ] A^{fi} (\Lambda ^{\prime} +1, \Lambda 
 ^{\prime} )  \nonumber \\ 
- { z \Lambda^{\prime} \over r \rho } A^{fi} (\Lambda ^{\prime} +1, 
\Lambda ^{\prime} ) \biggr \} d \> {\rm cos} \theta ,  
\label{jc23}
\end{eqnarray} 
where the factors $U_{-1} (f,i)$ are defined in (\ref{afi}). The
notations $\partial \rho_i$ and $\partial z_i$ mean that only the 
wave functions of the initial state in $A^{fi}$ must be
differentiated.  

The transverse E2 magnetization currents give rise to the following 
q.p. matrix elements:
\begin{eqnarray}
{\cal J}^S _{211} (fi,r) = i g^s \mu_N U_{-1} (f,i) {1 \over 8} 
\sqrt{3 \over \pi}  \int \biggl \{ {z \over r} \Bigl [ 
\Bigl ( {\partial \over \partial \rho } - {1 \over \rho } \Bigr ) 
\bigl [ A^{fi} (\Lambda +1, \Lambda ) - A^{fi} (\Lambda +2, 
\Lambda +1 ) \bigr ] \nonumber \\
+ 2  {\partial \over \partial z } A^{fi} (\Lambda  +1, \Lambda +1) 
\Bigr ] 
\nonumber \\
+ {\rho \over r} \Bigl [ {\partial \over \partial \rho } \bigl [ 
A^{fi} (\Lambda  +1, \Lambda +1) - A^{fi} (\Lambda  +2, \Lambda ) 
\bigr ] + { 2 \over \rho } A^{fi} (\Lambda  +2, \Lambda ) 
\Bigr ] \biggr \} d \> {\rm cos} \theta , \label{js21}
\end{eqnarray}
\begin{eqnarray}
{\cal J}^S _{231} (fi,r) = {i g^s \mu_N \over 4 \sqrt{2 \pi} } 
U_{-1} (f,i) \int \biggl \{  {z \over r} {\partial \over \partial 
\rho } \bigl [ A^{fi} (\Lambda +1, \Lambda ) - A^{fi} (\Lambda +2, 
\Lambda +1) \bigr ] \nonumber \\
+ { (z^2 - 4 \rho ^2 ) \over r^3 }  \Bigl [ \rho {\partial \over 
\partial \rho } \bigl [ A^{fi} (\Lambda +2, \Lambda ) 
- A^{fi} (\Lambda +1, \Lambda +1) \bigr ] \nonumber \\ 
- 2 A^{fi} (\Lambda  +2, \Lambda ) 
+  { z \over \rho }  \bigl [ A^{fi} (\Lambda  +1, \Lambda ) 
- A^{fi} (\Lambda  +2, \Lambda +1) \bigr ] \Bigl ] \nonumber \\
- { z \over r^3 } \Bigl [ \bigl ( 2z^2 - 3 \rho ^2 \bigr )  
 {\partial \over \partial z } A^{fi} (\Lambda  +1, \Lambda +1)  
 + 5 \rho ^2  {\partial \over \partial z }
 A^{fi} (\Lambda +2, \Lambda ) \Bigr ]  \biggr \} 
d \> {\rm cos} \theta , \label{js23}
\end{eqnarray}
where $g^s$ are the spin gyromagnetic ratios (\ref{grat}). All
differentiations in (\ref{jc21})-(\ref{js23}) are performed
analytically \cite{fanota} using recurrency relations for Hermite and
Laguerre polynomials.  The numerical integration in (\ref{rhome}) and 
(\ref{jc21})-(\ref{js23}) extends over $0 \le \cos \theta \le 1$.

\newpage 

\begin{figure}
\caption{Upper plot: DWBA cross section of the strongest M1 excitation
with $I^{\pi}K =1^+1$ and $E_x^{\rm th} = 2.9$ MeV in $^{156}$Gd
plotted versus the incident electron energy $E_i$ and the scattering
angle $\theta$.  Lower plot: DWBA cross section of the accompanying
E2 excitation with $I^{\pi}K =2^+1$. The $(e,e^{\prime})$ cross
sections are obtained from the RPA transition densities
(\protect\ref{rhod}), (\protect\ref{jd}). The thick contour lines 
indicate orders of magnitude, while the dashed contour lines are drawn 
at 5 times the next lower order of magnitude.}
\label{fig1}
\end{figure}

\begin{figure}
\caption{The same as in Fig.\ \protect\ref{fig1} but the DWBA M1 and
E2 cross sections are plotted versus the effective momentum transfer
$q_{\rm eff}$ (\protect\ref{qeff}) instead of $E_i$.}
\label{fig2}
\end{figure}

\begin{figure}
\caption{PWBA E2 cross section of the same excitation as in Fig.\
\protect\ref{fig1} plotted versus the incident electron energy $E_i$
for small scattering angles $\theta = 0.1^{\circ}, \; 0.5^{\circ}, \;
1^{\circ}$ and 5$^{\circ}$. The total cross section (continuous
curve) is a sum of the longitudinal C2 (dashed curve) and transverse
E2 (dotted curve) contributions.}
\label{fig3}
\end{figure}

\begin{figure}
\caption{Cross sections for $\theta = 70.53^{\circ}$, where $V_T = 1$
from Eq. (\protect\ref{vlt}), so that each PWBA cross section is
proportional to its form factor with the same coefficient.  Upper two
plots: PWBA cross sections versus momentum transfer $q$
(\protect\ref{qeff}).  Top plot: convection M1 (dashed curve) and
total M1 (convection plus spin, continuous curve). Middle plot:
longitudinal C2 (continuous curve) and transverse E2 (dashed curve).
Bottom plot:  DWBA cross sections versus $q_{\rm eff}$
(\protect\ref{qeff}); orbital M1 (long-dashed curve), total M1
(convection plus spin, continuous curve), and total E2 (longitudinal
plus transverse, short dashed curve).}
\label{fig4}
\end{figure}

\begin{figure}
\caption{Comparison between M1 (continuous curves) and E2 (dashed
curves) DWBA cross sections plotted versus the scattering angle
$\theta$. These are sections from the plots in Fig.
\protect\ref{fig1} corresponding to different incident energies
$E_i$, listed in each plot.} 
\label{fig5}
\end{figure}

\begin{figure}
\caption{M1 (upper plot) and E2 (lower plot) PWBA cross sections for
$\theta = 179.5^{\circ}$ versus momentum transfer $q$. Decomposition
and notations as in the upper two plots of Fig. \protect\ref{fig4},
respectively.}
\label{fig6}
\end{figure}

\begin{figure}
\caption{Comparison between M1 (continuous curves) and E2 (dashed
curves) DWBA cross sections plotted versus the effective momentum
transfer $q_{\rm eff}$ (\protect\ref{qeff}). These are sections from
Fig. \protect\ref{fig2} for different scattering angles $\theta$, 
listed in each plot.}
\label{fig7}
\end{figure}

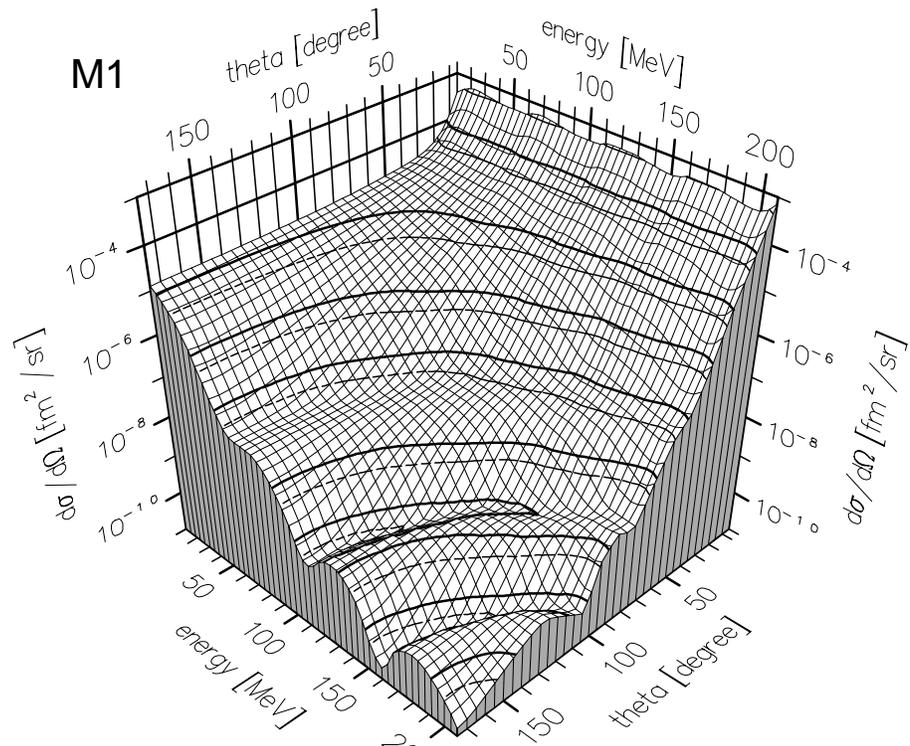
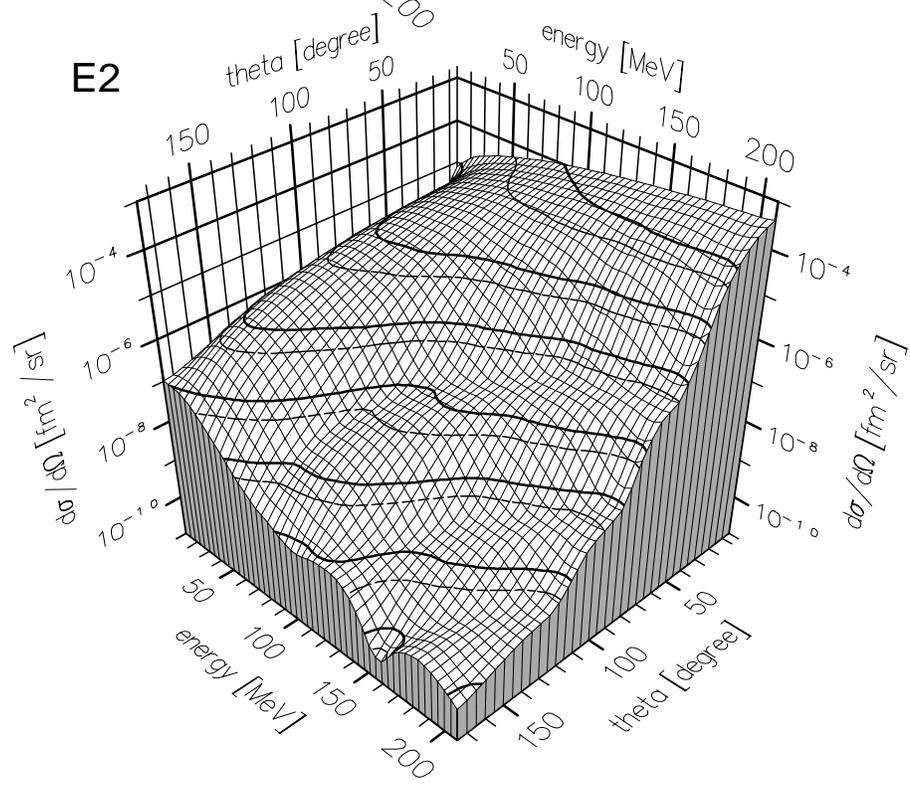
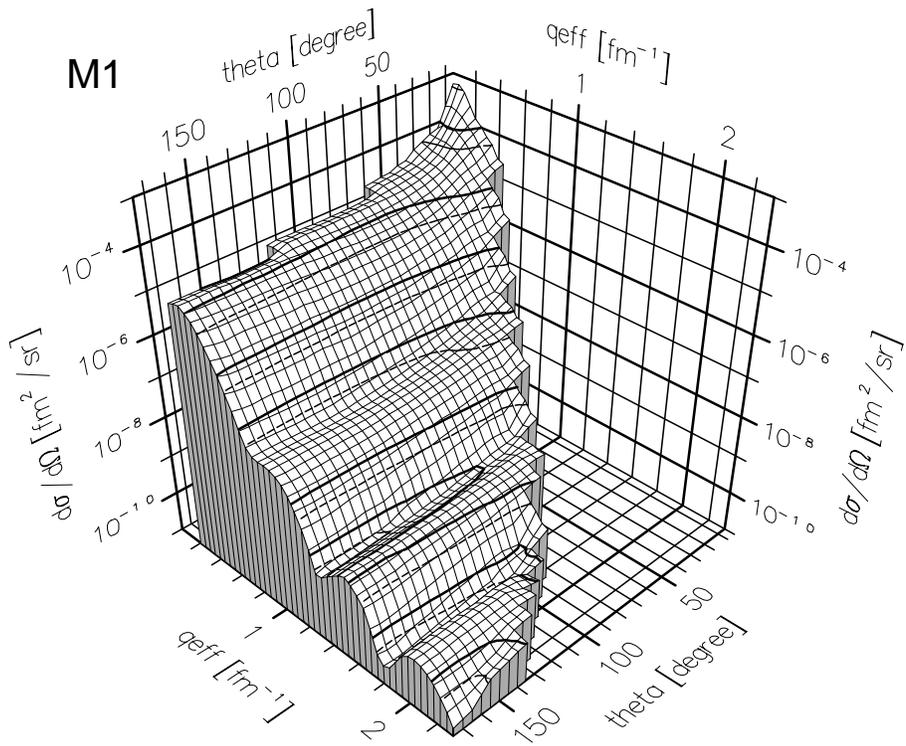
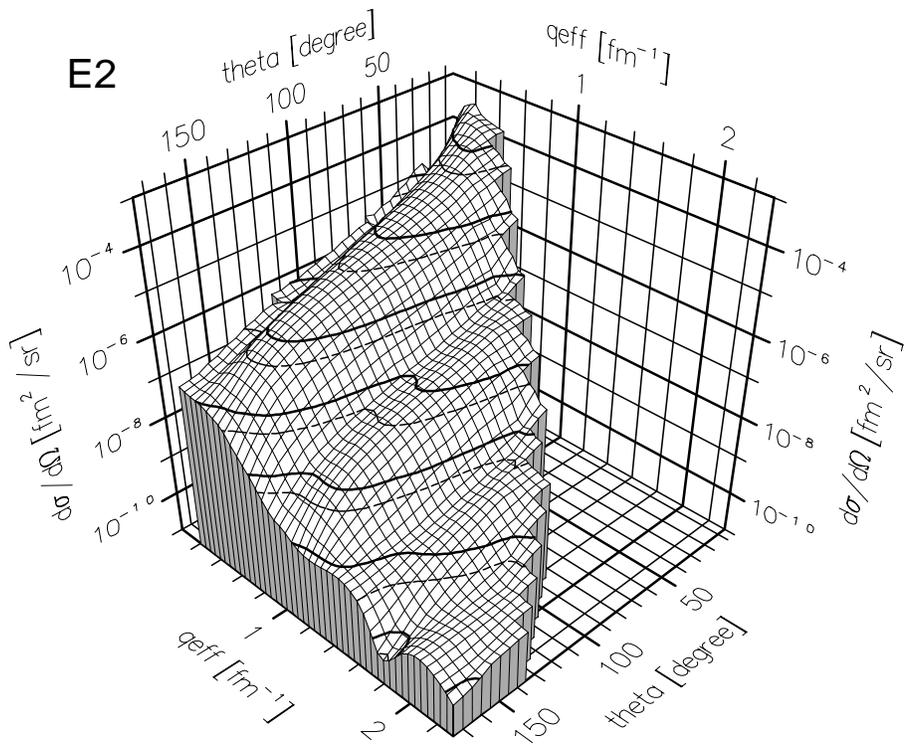
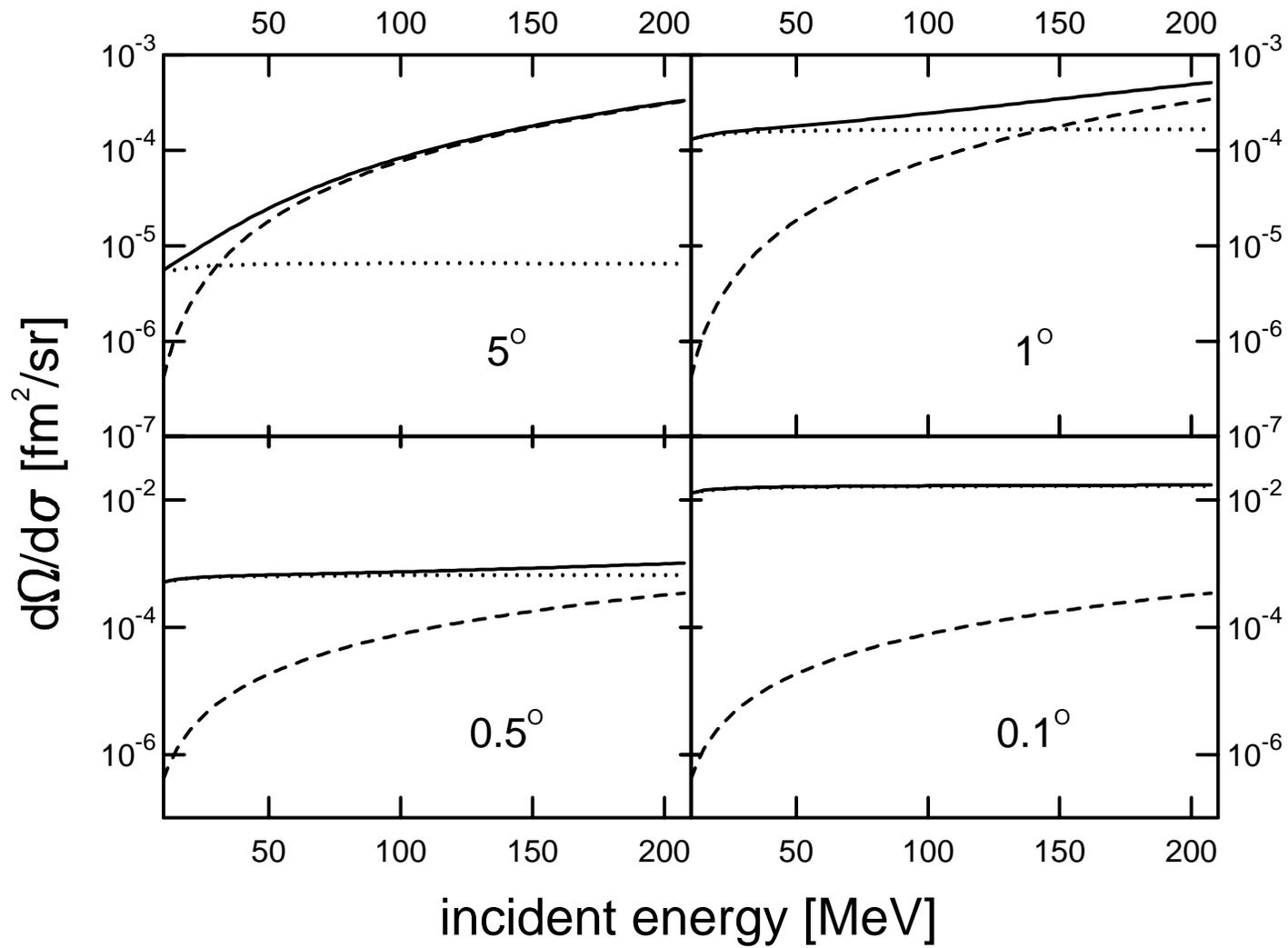
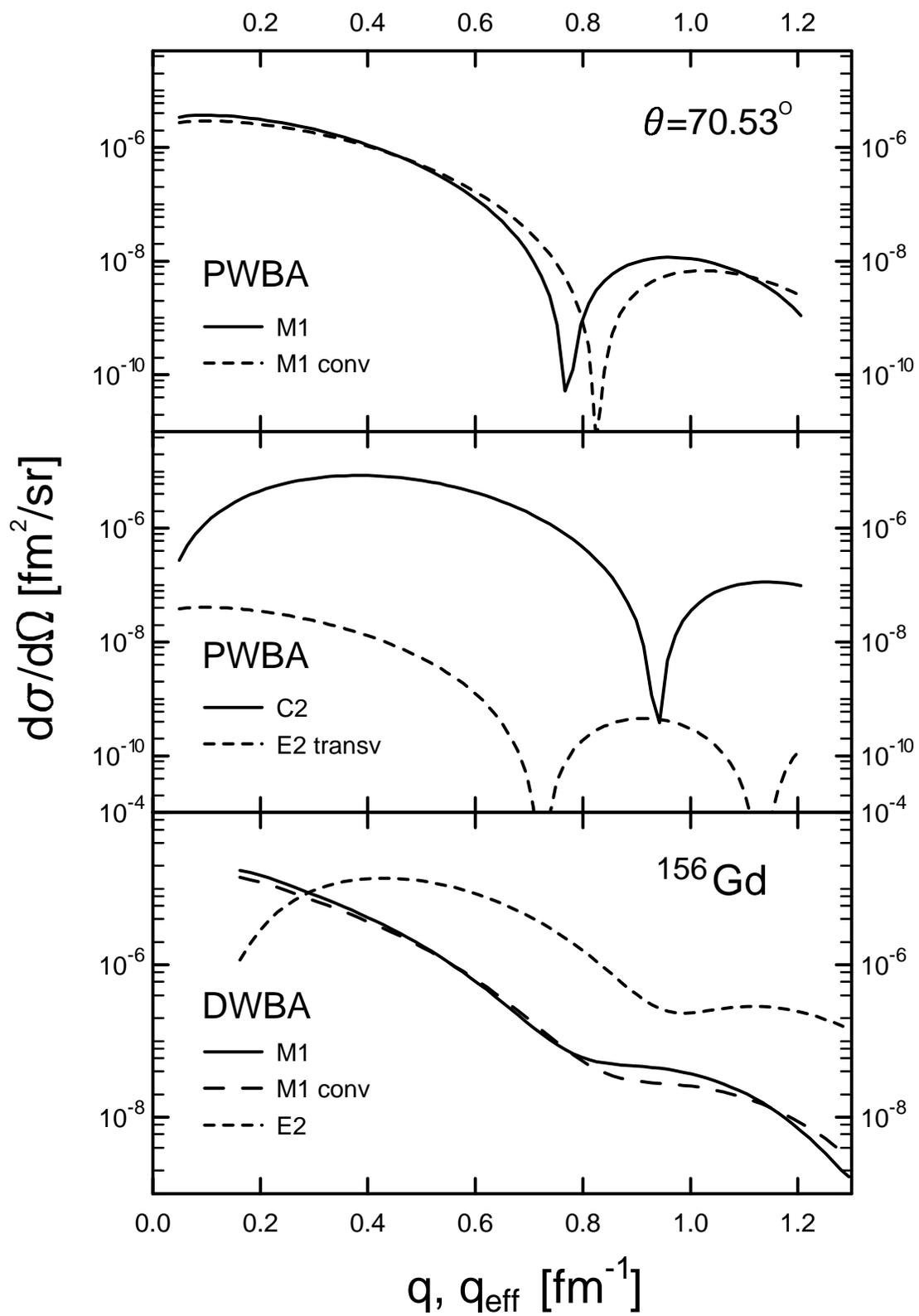
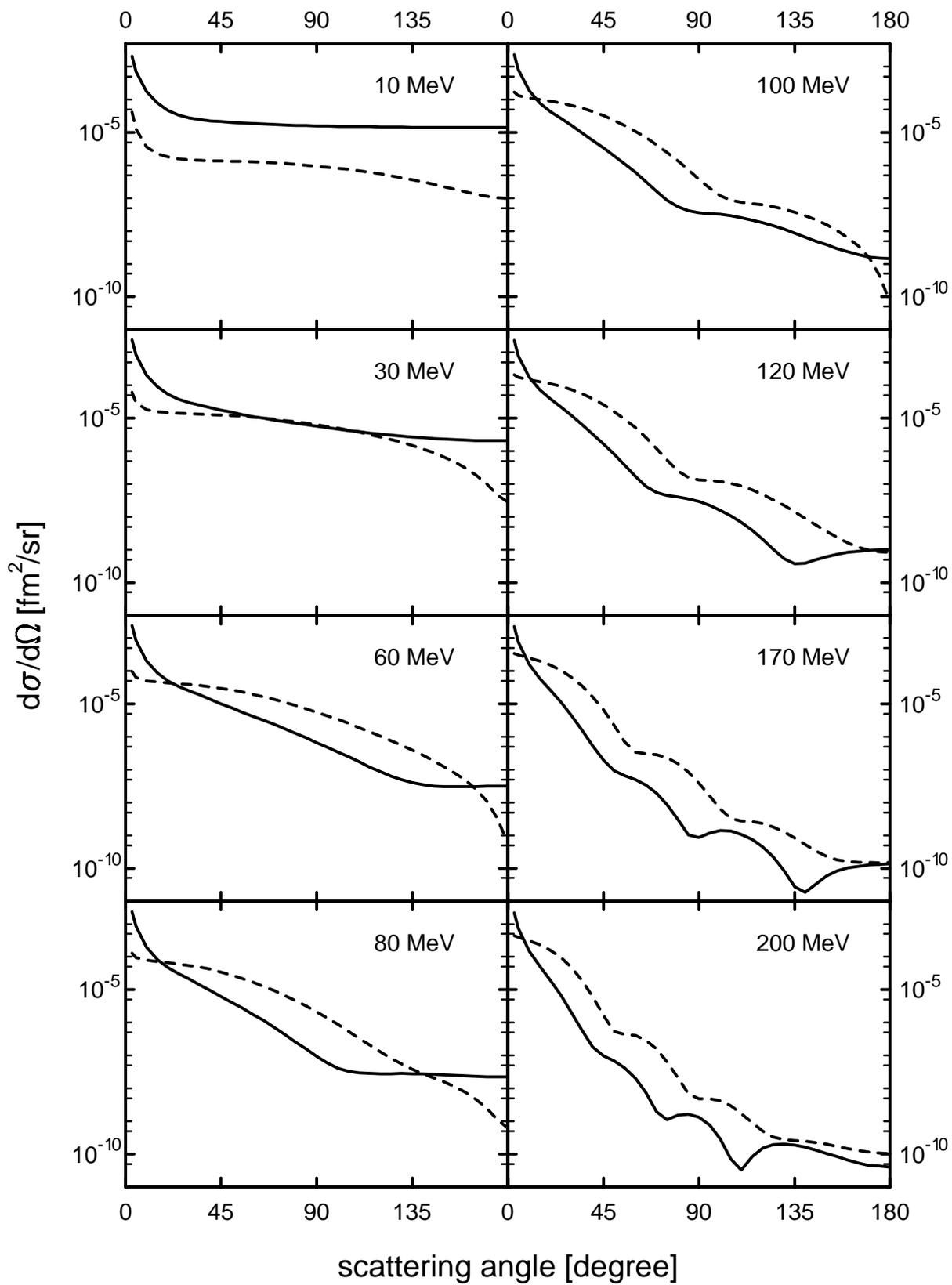
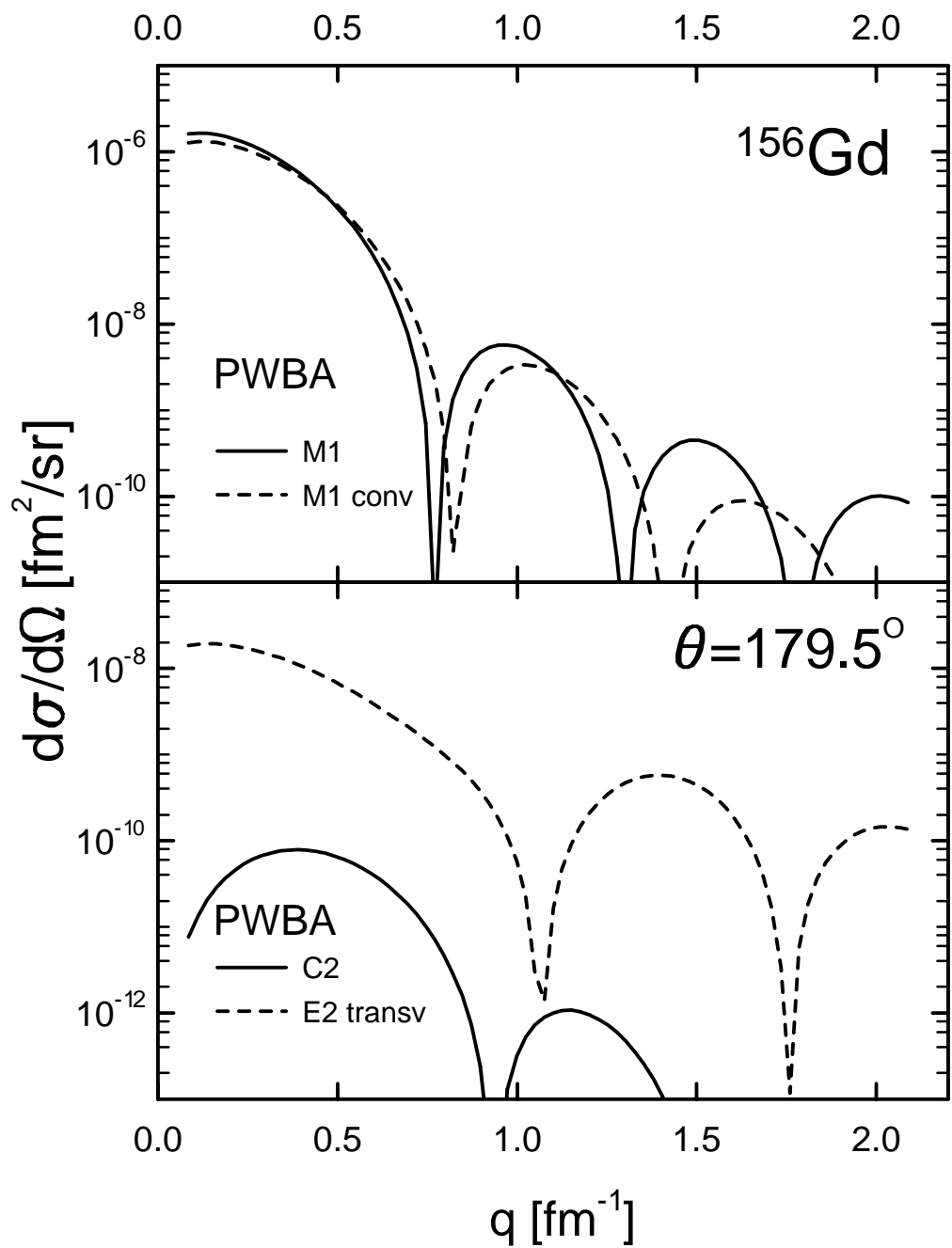
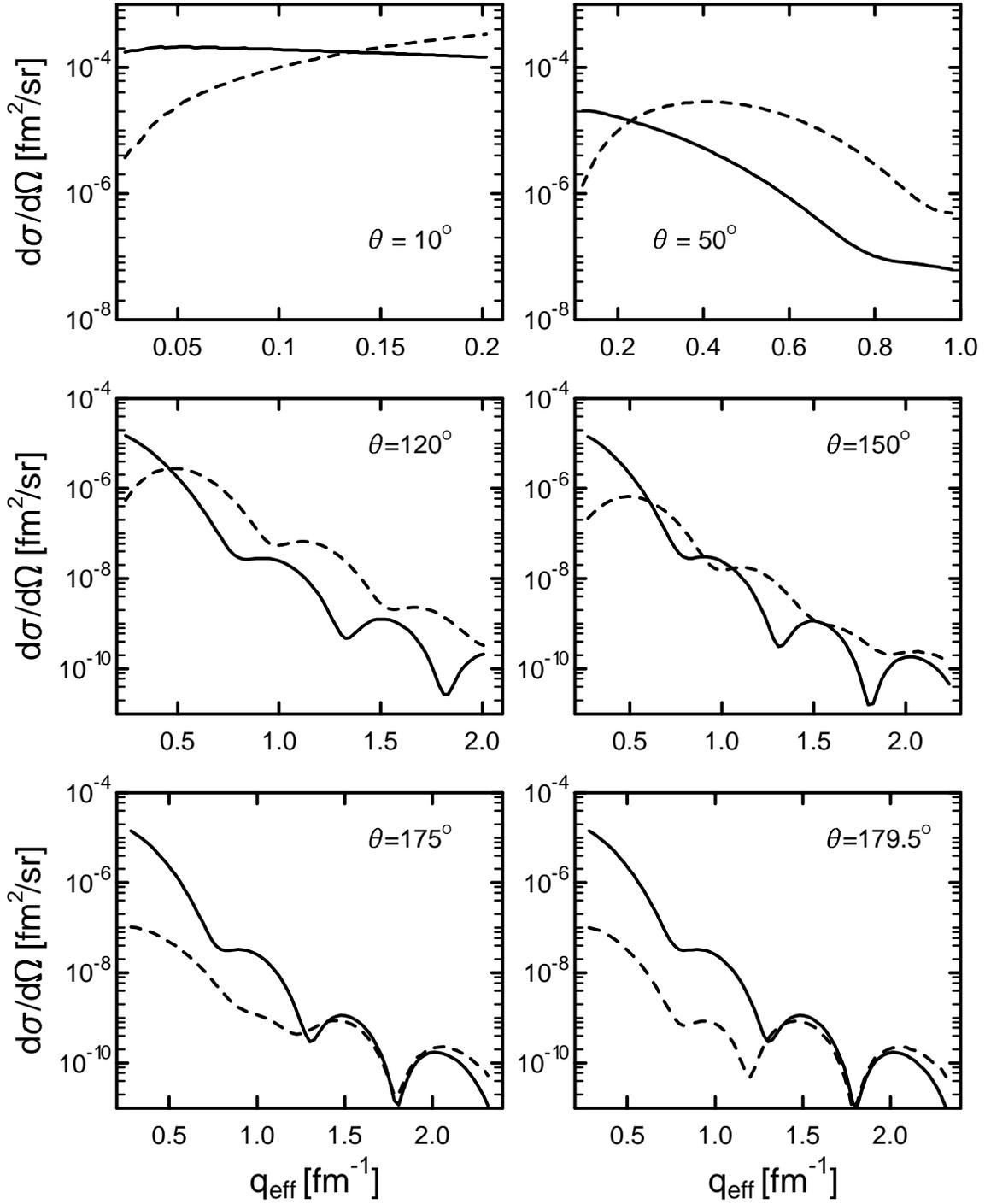
\end{document}